\documentclass{LMCS}

\usepackage{xspace}
\usepackage{amsmath}
\usepackage{amssymb} 
\usepackage{enumerate}
\usepackage{hyperref}

\newcounter{ncomm}

\newcommand{\CRE}{\textrm{CRE}}
\newcommand{\pcc}{T}

\newcommand{\gopcc}[3]{\mmgo_{\!#2}  #1. #3}

\newcommand{\allows}[3]{#1 \vdash^{#2}_{#3}}

\newcommand{\node}[3]{#1 [\![ \: #2 \: |\!\rangle\: #3 \: ]\!]}

\newcommand{\loc}{{\tt loc}}
\newcommand{\lgood}{{\tt lgood}}
\newcommand{\lbad}{{\tt lbad}}

\newcommand{\dpisubt}{<:}

\newcommand{\pic}{$\pi$--calculus}

\newcommand{\emptynet}{{\bf 0}}

\newcommand{\define}{
 \mbox{$\,\mathop{\hbox to \wd0{\,\,=\,\,}}\limits^{\raisebox{-.13ex}{\hbox{\tiny $\triangle $}}}\,$}
}

\newcommand{\mmif}{{\bf if\ }}

\newcommand{\mmthen}{{\bf \ then\ }}
\newcommand{\mmelse}{{\bf \ else\ }}
\newcommand{\mmgo}{{\bf go\,}}

\newcommand{\mmnil}{{\bf nil}}

\newcommand{\mmin}{{\bf in}}
\newcommand{\mmlet}{{\bf let}}

\makeatletter
\def \rightarrowfill{\m@th\mathord{\smash-}\mkern-6mu%
  \cleaders\hbox{$\mkern-2mu\mathord{\smash-}\mkern-2mu$}\hfill
  \mkern-6mu\mathord\rightarrow}
\makeatother
\def \overstackrel#1#2{\mathrel{\mathop{#1}\limits^{#2}}}
\newcommand{\redar}[1]{\mbox{ $\overstackrel \rightarrow{#1}$ }}
\newcommand{\red}[1]{\mbox{ $\overstackrel{\rightarrow}{#1}$ }}

\newdimen\proofrulebreadth \proofrulebreadth=.04em
\newdimen\proofdotseparation \proofdotseparation=1.25ex
\newdimen\proofrulebaseline \proofrulebaseline=2ex
\newcount\proofdotnumber \proofdotnumber=3
\let\then\relax
\def\hfi{\hskip0pt plus.0001fil}
\mathchardef\squigto="3A3B
\newif\ifinsideprooftree\insideprooftreefalse
\newif\ifonleftofproofrule\onleftofproofrulefalse
\newif\ifproofdots\proofdotsfalse
\newif\ifdoubleproof\doubleprooffalse
\let\wereinproofbit\relax
\newdimen\shortenproofleft
\newdimen\shortenproofright
\newdimen\proofbelowshift
\newbox\proofabove
\newbox\proofbelow
\newbox\proofrulename
\def\shiftproofbelow{\let\next\relax\afterassignment\setshiftproofbelow\dimen0 }
\def\shiftproofbelowneg{\def\next{\multiply\dimen0 by-1 }%
\afterassignment\setshiftproofbelow\dimen0 }
\def\setshiftproofbelow{\next\proofbelowshift=\dimen0 }
\def\setproofrulebreadth{\proofrulebreadth}

\def\prooftree{
\ifnum  \lastpenalty=1 \then   \unpenalty \else
\onleftofproofrulefalse \fi
\ifonleftofproofrule \else   \ifinsideprooftree
        \then   \hskip.5em plus1fil
        \fi
\fi
\bgroup
\setbox\proofbelow=\hbox{}\setbox\proofrulename=\hbox{}%
\let\justifies\proofover\let\leadsto\proofoverdots\let\Justifies\proofoverdbl
\let\using\proofusing\let\[\prooftree
\ifinsideprooftree\let\]\endprooftree\fi
\proofdotsfalse\doubleprooffalse
\let\thickness\setproofrulebreadth
\let\shiftright\shiftproofbelow \let\shift\shiftproofbelow
\let\shiftleft\shiftproofbelowneg
\let\ifwasinsideprooftree\ifinsideprooftree
\insideprooftreetrue
\setbox\proofabove=\hbox\bgroup$\displaystyle 
\let\wereinproofbit\prooftree
\shortenproofleft=0pt \shortenproofright=0pt \proofbelowshift=0pt
\onleftofproofruletrue\penalty1 }

\def\eproofbit{
\ifx    \wereinproofbit\prooftree \then   \ifcase \lastpenalty
        \then   \shortenproofright=0pt  
        \or     \unpenalty\hfil         
        \or     \unpenalty\unskip       
        \else   \shortenproofright=0pt  
        \fi
\fi
\global\dimen0=\shortenproofleft \global\dimen1=\shortenproofright
\global\dimen2=\proofrulebreadth \global\dimen3=\proofbelowshift
\global\dimen4=\proofdotseparation
$\egroup  
\shortenproofleft=\dimen0 \shortenproofright=\dimen1
\proofrulebreadth=\dimen2 \proofbelowshift=\dimen3
\proofdotseparation=\dimen4
}

\def\proofover{
\eproofbit 
\setbox\proofbelow=\hbox\bgroup 
\let\wereinproofbit\proofover
$\displaystyle
}%
\def\proofoverdbl{
\eproofbit 
\doubleprooftrue
\setbox\proofbelow=\hbox\bgroup 
\let\wereinproofbit\proofoverdbl
$\displaystyle

}%
\def\proofoverdots{
\eproofbit 
\proofdotstrue
\setbox\proofbelow=\hbox\bgroup 
\let\wereinproofbit\proofoverdots

$\displaystyle
}%
\def\proofusing{
\eproofbit 
\setbox\proofrulename=\hbox\bgroup 
\let\wereinproofbit\proofusing
\kern0.3em$ }

\def\endprooftree{
\eproofbit 
  \dimen5 =0pt
\dimen0=\wd\proofabove \advance\dimen0-\shortenproofleft
\advance\dimen0-\shortenproofright
\dimen1=.5\dimen0 \advance\dimen1-.5\wd\proofbelow \dimen4=\dimen1
\advance\dimen1\proofbelowshift \advance\dimen4-\proofbelowshift
\ifdim  \dimen1<0pt \then   \advance\shortenproofleft\dimen1
        \advance\dimen0-\dimen1
        \dimen1=0pt
        \ifdim  \shortenproofleft<0pt
        \then   \setbox\proofabove=\hbox{%
                        \kern-\shortenproofleft\unhbox\proofabove}%
                \shortenproofleft=0pt
        \fi
\fi
\ifdim  \dimen4<0pt \then   \advance\shortenproofright\dimen4
        \advance\dimen0-\dimen4
        \dimen4=0pt
\fi
\ifdim  \shortenproofright<\wd\proofrulename \then
\shortenproofright=\wd\proofrulename \fi
\dimen2=\shortenproofleft \advance\dimen2 by\dimen1
\dimen3=\shortenproofright\advance\dimen3 by\dimen4
\ifproofdots \then
        \dimen6=\shortenproofleft \advance\dimen6 .5\dimen0
        \setbox1=\vbox to\proofdotseparation{\vss\hbox{$\cdot$}\vss}
        \setbox0=\hbox{%
                \kern\dimen6
                $\vcenter to\proofdotnumber\proofdotseparation
                        {\leaders\box1\vfill}$%
                \unhbox\proofrulename}%
\else   \dimen6=\fontdimen22\the\textfont2 
        \dimen7=\dimen6
        \advance\dimen6by.5\proofrulebreadth
        \advance\dimen7by-.5\proofrulebreadth
        \setbox0=\hbox{%
                \kern\shortenproofleft
                \ifdoubleproof
                \then   \hbox to\dimen0{%
                        $\mathsurround0pt\mathord=\mkern-6mu%
                        \cleaders\hbox{$\mkern-2mu=\mkern-2mu$}\hfill
                        \mkern-6mu\mathord=$}%
                \else   \vrule height\dimen6 depth-\dimen7 width\dimen0
                \fi
                \unhbox\proofrulename}%
        \ht0=\dimen6 \dp0=-\dimen7
\fi
\let\doll\relax
\ifwasinsideprooftree \then   \let\VBOX\vbox \else
\ifmmode\else$\let\doll=$\fi
        \let\VBOX\vcenter
\fi
\VBOX   {\baselineskip\proofrulebaseline \lineskip.2ex
        \expandafter\lineskiplimit\ifproofdots0ex\else-0.6ex\fi
        \hbox   spread\dimen5   {\hfi\unhbox\proofabove\hfi}%
        \hbox{\box0}%
        \hbox   {\kern\dimen2 \box\proofbelow}}\doll%
\global\dimen2=\dimen2 \global\dimen3=\dimen3
\egroup 
\ifonleftofproofrule \then   \shortenproofleft=\dimen2 \fi
\shortenproofright=\dimen3
\onleftofproofrulefalse \ifinsideprooftree \then   \hskip.5em plus
1fil \penalty2 \fi }

\makeatletter
\DeclareSymbolFont{smallcaps}{\encodingdefault}{\rmdefault}{m}{sc}
\DeclareSymbolFontAlphabet\mathsc{smallcaps}

\newcommand{\synf}[1]{ \ensuremath{\mathsc{#1}}} 
\newcommand{\locf}[1]{ \ensuremath{\mathsc{#1}}} 
\newcommand{\actf}[1]{ \ensuremath{\mathtt{#1}}} 

\newcommand{\Cinfo}{\actf{info}\xspace}
\newcommand{\Crequest}{\actf{req}\xspace}
\newcommand{\Ctake}{\actf{take}\xspace}
\newcommand{\Cgive}{\actf{give}\xspace}
\newcommand{\Cbob}{\locf{bob}\xspace}
\newcommand{\Calice}{\locf{alice}\xspace}

\newcommand{\Chome}{\locf{home}\xspace}
\newcommand{\Csecure}{\locf{secure}\xspace}

\newcommand{\Cmailserv}{\locf{mail\_serv}\xspace}
\newcommand{\Cspam}{\locf{spam}\xspace}
\newcommand{\Clicence}{\locf{licence\_serv}\xspace}
\newcommand{\Cusr}{\actf{usr}\xspace}
\newcommand{\Cpwd}{\actf{pwd}\xspace}
\newcommand{\Clist}{\actf{list}\xspace}
\newcommand{\Csend}{\actf{send}\xspace}
\newcommand{\Cretr}{\actf{retr}\xspace}
\newcommand{\Cdel}{\actf{del}\xspace}
\newcommand{\Creset}{\actf{reset}\xspace}
\newcommand{\Cquit}{\actf{quit}\xspace}
\newcommand{\Clock}{\actf{lock}\xspace}
\newcommand{\Cunlock}{\actf{unlock}\xspace}
\newcommand{\Csecret}{\actf{secret}\xspace}
\newcommand{\Cgetlic}{\actf{get\_licence}\xspace}

\newcommand{\Act}{\synf{Act}\xspace}

\newcommand{\Locs}{\synf{Loc}\xspace}

\newcommand{\Sset}[1]{\mathop{{\tt act}}(#1)}

\newcommand{\memb}{M}
\renewcommand{\loc}{{\tt unknown}}
\renewcommand{\lgood}{{\tt good}}
\renewcommand{\lbad}{{\tt bad}}
\renewcommand{\pcc}{{\pol{T}}}
\newcommand{\redstar}[1]{ \mbox{$\;\;\overstackrel{\rightarrowfill^*}{#1}\;\;$} }
\newcommand{\pol}[1]{\mathsf{#1}}

\newcommand{\sset}[1]{\{ {#1}  \}  } 
\newcommand{\types}{\vdash}
\newcommand{\infers}{\Vdash}
\newcommand{\tok}{\mathbf{\scriptstyle ok}}
\newcommand{\enforces}{\,\mathbin{{\tt enforces}}\,}
\newcommand{\join}{\mathbin{\sqcup}}
\newcommand{\Ptype}[1]{\pol{pol}(#1)}
\newcommand{\giveme}{\succ}
\DeclareSymbolFont{smallcaps}{\encodingdefault}{\rmdefault}{m}{sc}
\DeclareSymbolFontAlphabet\mathsc{smallcaps}
\newcommand{\Rulef}[1]{{\mathsc{#1}}}   
\newcommand{\Rred}[1]{\ensuremath{\Rulef{(r\textrm{-}#1)}}}
\newcommand{\Rtype}[1]{\ensuremath{\Rulef{(tc\textrm{-}#1)}}}
\newcommand{\Rinf}[1]{\ensuremath{\Rulef{(ti\textrm{-}#1)}}}
\newcommand{\Rstype}[1]{\ensuremath{\Rulef{(wf\textrm{-}#1)}}}
\newcommand{\Rlts}[1]{\ensuremath{\Rulef{(lts\textrm{-}#1)}}}
\newcommand{\mnowidth}[2][l]{\makebox[0cm][#1]{$#2$}}%

\RequirePackage{ifthen}   
\newcommand{\ifemptyelse}[3]{\ifx\@@bullshit#1\@@bullshit#2\else#3\fi}
\newcommand{\ifnotempty}[2]{\ifx\@@bullshit#1\@@bullshit\else#2\fi}
\newenvironment{scope}
  {\bgroup\ignorespaces}
  {\egroup\global\@ignoretrue}
\newbox\@topbox
\newbox\@botbox
\newcommand{\linfer}[3][{}]{%
  \ifemptyelse{#1}{
    \setbox\@topbox\hbox{%
      \renewcommand{\arraystretch}{1}%
      $\begin{array}{l}#2\end{array}$}%
    }{
    \ifemptyelse{#2}{
      \setbox\@topbox\hbox{%
        \renewcommand{\arraystretch}{1}%
        $\begin{array}{l}\mnowidth[l]{\scriptstyle#1}\\\ \end{array}$}%
      }{
      \setbox\@topbox\hbox{%
        \renewcommand{\arraystretch}{1}%
        $\begin{array}{l}\mnowidth[l]{\scriptstyle#1}\\#2\end{array}$}%
      }
    }
  \setbox\@botbox\hbox{%
    \def\tilde##1{\widetilde{##1}}%
    \renewcommand{\arraystretch}{1}%
    $\begin{array}{l}#3\end{array}$}%
  \ifdim\wd\@botbox<\wd\@topbox
    \setbox\@botbox\hbox to \wd\@topbox{\box\@botbox\hfil}
  \else
    \setbox\@topbox\hbox to \wd\@botbox{\box\@topbox\hfil}
  \fi
  \frac{\box\@topbox}{\box\@botbox}
  }
\newcommand{\linferSIDE}[4][{}]{%
  \ifemptyelse{#1}{
    \setbox\@topbox\hbox{%
      \renewcommand{\arraystretch}{1}%
      $\begin{array}{l}#2\end{array}$}%
    }{
    \ifemptyelse{#2}{
      \setbox\@topbox\hbox{%
        \renewcommand{\arraystretch}{1}%
        $\begin{array}{l}\mnowidth[l]{\scriptstyle#1}\\\ \end{array}$}%
      }{
      \setbox\@topbox\hbox{%
        \renewcommand{\arraystretch}{1}%
        $\begin{array}{l}\mnowidth[l]{\scriptstyle#1}\\#2\end{array}$}%
      }
    }
  \setbox\@botbox\hbox{%
    \def\tilde##1{\widetilde{##1}}%
    \renewcommand{\arraystretch}{1}%
    $\begin{array}{l}#3\end{array}$}%
  \ifdim\wd\@botbox<\wd\@topbox
    \setbox\@botbox\hbox to \wd\@topbox{\box\@botbox\hfil}
  \else
    \setbox\@topbox\hbox to \wd\@botbox{\box\@topbox\hfil}
  \fi
  \frac{\box\@topbox}{\box\@botbox}
  \hbox{\;\footnotesize $
    \begin{scope}%
      \renewcommand{\arraystretch}{1}%
      \begin{array}[c]{l}%
        #4%
      \end{array}%
    \end{scope}%
    $}%
}

\newcommand{\slinfer}[2][{}]{\renewcommand{\arraystretch}{1}%
 \begin{array}{l}{\scriptstyle #1}\\#2\end{array}}

\newenvironment{myfigure}[1][tbp]{\begin{figure}[#1]%
                                     \hrule height1pt\vspace{2ex}}
                                {\vspace{2ex}\hrule height1pt\vspace{3ex}\vspace*{-.7cm}
\end{figure}}

\def\fps@figure{tp}      
\def\fps@table{tp}
\newtheorem{defii}{defi}[section]
\newtheorem{exx}{exa}[section]
\newcommand{\EndProofBox}{\null\hfill$\blacksquare$}
\global\let\EndProof\EndProofBox
\newcommand{\boxHere}{\global\let\EndProof\empty\EndProofBox}

\makeatother

\def\doi{1 (3:2) 2005}
\lmcsheading%
{\doi}
{331--353}
{}
{}
{Feb.~\phantom{0}5, 2005}
{Dec.~20, 2005}
{}

\begin{document}

\title[Security Policies as Membranes in Systems for Global Computing]{Security Policies as Membranes in Systems for Global Computing}

\author[D.~Gorla]{Daniele Gorla\rsuper a}
\address{{\lsuper a}Dip. di Informatica, Univ. di Roma ``La Sapienza"}
\email{gorla@di.uniroma1.it}
\thanks{{\lsuper a}This work has been mostly carried on while the first author
was at the Dept. of Informatics, Univ. of Sussex, with a Marie Curie Fellowship.}

\author[M.~Hennessy]{Matthew Hennessy\rsuper b}
\address{{\lsuper b}Dept. of Informatics, Univ. of Sussex}
\email{matthewh@sussexx.ac.uk}
\thanks{{\lsuper{a, b, c}}The authors would like to acknowledge the support of the EC
  Global Computing projects Mikado and Myths.}

\author[V.~Sassone]{Vladimiro Sassone\rsuper c}
\address{{\lsuper c}Dept. of Informatics, Univ. of Sussex}
\email{vs@sussex.ac.uk}

\begin{abstract}
%
 We propose a simple global computing framework, whose main concern is
 code migration. Systems are structured in sites, and each site is
 divided into  two parts: a computing body, and a {\em membrane} which
 regulates the interactions between the computing body and the
 external environment.
 More precisely, membranes are filters which control access to the
 associated site,  and they also rely  on the well-established notion of
 {\em trust} between sites. 
 We develop a basic theory to express and enforce security
 policies via membranes. Initially, these only control the actions
 incoming agents intend to perform locally.  
 We then adapt the basic theory to encompass more sophisticated
 policies, where the number of actions an agent wants to perform, and
 also their order, are considered. 
\end{abstract}

\maketitle

\section{Introduction}

 Computing is increasingly characterised by the global
 scale of applications and the ubiquity of interactions between mobile
 components. Among the main features of the forthcoming ``global
 ubiquitous computing'' paradigm we list 
 \textit{distribution} and \textit{location awarness}, 
   whereby code located at specific sites acts appropriately to
   local parameters and circumstances, that is, it is    ``context-aware'';
 \textit{mobility}, whereby  code is dispatched from site to site to
   increase flexibility and expressivity;
 \textit{openness}, reflecting the nature of global networks and
   embodying the permeating hypothesis of localised, partial knowledge
   of the execution environment. 
 Such systems present enormous difficulties, both technical and
 conceptual, and are currently more at the stage of exciting future
 prospectives than  that of established of engineering practice. 
 Two concerns, however, appear to clearly have a ever-reaching import:
 \textit{security} and \textit{mobility control}, arising respectively
 from openness and from massive code and resource migrations.
 They are the focus of the present paper.

 We aim at classifying mobile components according to their
 {behaviour},  and at empowering sites with control
 capabilities which allow them to deny access to those agents 
 whose behaviour does not conform to the site's \textit{policy}. 
 We see every site of a system 
$$
\node k \memb P
$$
as an entity named $k$ and structured in two layers: a 
 \textit{computing body} $P$, where programs run their code --
  possibly accessing local resources offered by the site -- and a 
 \textit{membrane} $\memb$, which regulates the interactions between the
 computing body and the external environment. 
 An agent $P$ wishing to enter a site $l$ must be verified by the
 membrane before it is given a chance to execute in $l$. If the
 preliminary check succeeds, the agent is allowed to execute, otherwise
 it is rejected. In other words, a membrane implements the policy each
 site wants to enforce locally, by ruling on the requests of access of
 the incoming agents. This can be easily expressed by a migration
 rule of the form: 
\[
  \node k {\memb^k} {\gopcc l {} P\mid Q} \parallel 
  \node l {\memb^l} R  \red{} 
  \node k {\memb^k} {Q} \ \parallel \ \node l {\memb^l} {P\mid R}
 \qquad\text{if }\allows {\memb^l} k {} P
\]
The relevant parts here are $P$, the agent wishing to migrate from $k$
to $l$, and $l$, the receiving site, which needs to be satisfied that
$P$'s behaviour complies with its policy. The latter is expressed by
$l$'s membrane, $\memb^l$. The judgement $\allows {\memb^l} k {} P$
represents $l$ inspecting the incoming code to verify that it upholds
$\memb^l$.

 Observe that in the formulation above $\allows {\memb^l} k {} P$
 represents a runtime check of all incoming agents. Because of our
 fundamental assumption of openendedness, such kind of checks,
 undesirable as they might be, cannot be avoided. 
 In order to reduce their impact on systems performance, and to make the
 runtime semantics as efficient as possible, we adopt a strategy
which allows  for efficient agent verification. Precisely, we adopt an elementary
 notion of \textit{trust}, so that from the point of view of each $l$
 the set of sites is consistently partitioned between ``good,''
 ``bad,'' and ``unknown'' sites. Then, in a situation like the one in
 the rule above, we assume that $l$ will be willing to accept from a
 \textit{trusted} site $k$ a $k$-\textit{certified digest} $\pcc$ of
 $P$'s behaviour. We then modify the primitive $\mathbf{go}$ and the
 judgement $\vdash^k$ as in the refined migration rule:
 \[
  \node k {\memb^k} {\gopcc l \pcc P \ | \ Q} \parallel 
         \node l {\memb^l} R  \red{} 
  \node k {\memb^k} {Q} \ \parallel \ \node l {\memb^l} {P\,|\,R}
 \qquad\mbox{if }\allows {\memb^l} k \pcc P
\]
 The notable difference is in $\allows {\memb^l} k \pcc P$. Here, $l$
 verifies the entire code $P$ against $\memb^l$ \textit{only if} it
 does not trust $k$, the signer of $P$'s certificate $\pcc$. 
 Otherwise, it suffices for $l$ to match $\memb^l$ against the
 digest $\pcc$ carried by $\mathbf{go}$ together with $P$ from $k$, so
 effectively shifting work from $l$ to the originator of $P$.

 Our main concern in this paper is to put the focus on the machinery a
 membrane should implement to enforce \textit{different kinds} of
 policies. We first distill the simplest calculus which can
 conceivably convey our ideas and still support a non-trivial study. 
 It is important to remark that we are abstracting from agents' local
 computations. These can be expressed in any of several well-known
 models for concurrency, for example CCS \cite{milner:CCS} or the \pic\
 \cite{milner:communicating-mobile}. We are concerned, instead, with
 agents' migration from site to site: our main language mechanism
 is $\mathbf{go}$ rather than intra-site (i.e.\ local) communication.
 Using this language, we examine four notions of policy and show
 how they can be enforced by using membranes. We start with an
 amusingly simple policy which only lists allowed actions. We then
 move to count action occurrences and then to policies expressed by 
 \textit{deterministic finite automata}. 
 Note that such policies are only concerned with the behaviour of
 single agents, and do \textit{not} take into account 
 ``\textit{coalitional}''
 behaviours, whereby incoming agents -- apparently innocent -- join
 clusters of resident agents -- they too apparently innocent -- to
 perform cooperatively potentially harmful actions, or at least overrule
 the host site's policy. We call \textit{resident} those policies
 intended to be applied to the joint, composite behaviour of the
 agents contained at a site. We explore resident policies as our fourth
 and final notion of policy. In all the cases, the theory adapts
 smoothly; we only need to refine the information
 stored in the membrane and the inspection mechanisms.\medskip

\noindent{\bf Structure of the paper.}
 In Section~\ref{calculus} we define the calculus used in this paper,
 and start with the straightforward policy which only prescribes the
 actions an agent can perform when running in a site.
 In Section~\ref{entryPol}, we enhance the theory to control also how
 many (and not only which kind of) actions an agent wants to perform
 in a site, and their order of execution. 
 Finally, in Section~\ref{residentPol} we extend the theory to control
 the overall computation taking place at a site, and not only the
 behaviour of single agents. The paper concludes in
 Section~\ref{conclRelWork} where a comparison with related work is
 also given. The theoretical results are proved in
 Appendix~\ref{techProofs}. With respect to the extended abstract
 \cite{GHS04:fguc}, this paper contains more examples together with
 complete proofs.

\section{A Simple Calculus} 
\label{calculus}
In this section we describe a simple calculus for mobile agents, 
which may migrate between sites. Each site is guarded by a \emph{membrane}, 
whose task is to ensure that every agent accepted
at the site conforms to an \emph{entry policy}.

\subsection{The Syntax}
\ \\

The syntax is given in Figure~\ref{BasicSyntax} and assumes two pairwise disjoint sets: 
basic agent actions $\Act$, ranged over by $a,b,c,\cdots$,
and localities $\Locs$, ranged over by $l,k,h,\cdots$.
Agents are constructed using the standard action-prefixing,
parallel composition and replication operators from process calculi,
\cite{milner:CCS}. The one novel operator is that for migration,
$$ 
\gopcc l \pcc P 
$$
This agent seeks to migrate to site $l$ in order to execute the code $P$; 
moreover it promises to conform to the entry policy $\pcc$. 
In practical terms this might consist of a certification that the incoming 
code $P$ conforms to the policy $\pcc$, which the site $l$ has to decide whether 
or not to accept. 
In our framework, this certification is a policy that describes the (local)
behaviour of the agent; thus, in $\gopcc l \pcc P$, $\pcc$ will be called
the {\em digest} of $P$.

\begin{myfigure} 
{\hfill
\begin{tabular}{rcclr} 
{\it Basic Actions} & \multicolumn{4}{l}{$a,b,c, ... \in \Act$} \\
{\it Localities} & \multicolumn{4}{l}{$l,h,k, ...\in \Locs $} \\
&\\ {\it Agents} & $P,Q,R$ & $::=$ & \mmnil & nil agent\\ 
& & $|$ & $a.P$ & basic action \\
& & $|$ & $\gopcc l \pcc P$& migration\\
& & $|$ & $P \ |\  Q$ & composition \\
& & $|$ & $!P$ & replication\\
&\\{\it Systems} & $N$ & ::=  & $\emptynet$ & empty system\\
& & $|$ & $\node l \memb P$ & site\\
& & $|$ & $N_1 \parallel N_2$ & composition \\ 
\end{tabular} 
\hfill}
\caption{A Simple Calculus} 
\label{BasicSyntax} 
\end{myfigure}

A system consists of a finite set of sites running in parallel. A site takes the form 
\begin{displaymath} 
\node l \memb P
\vspace*{-.2cm}
\end{displaymath} 
where
\begin{enumerate}[$\bullet$] 
\item $l$ is the site name 
\item $P$ is the code currently running at $l$ 
\item $\memb$ is the membrane which implements the entry policy. 
\end{enumerate} 
For convenience we assume that site names are unique in systems. Thus, 
in a given system we can identify the membrane associated with the site named 
$l$ by $\memb^l$.
We start with a very simple kind of policy, which we will then progressively enhance.

\begin{defi}[Policies] 
\label{def:policies} 
A \emph{policy} is a finite subset of $\Act \cup \Locs$. 
For two policies $\pol{T}_1$ and $\pol{T}_2$, we write 
\begin{displaymath}
\pol{T}_1 \enforces \pol{T}_2
\end{displaymath} 
whenever $\pol{T}_1 \subseteq \pol{T}_2$.
\end{defi}

Intuitively an agent conforms to a policy $\pol{T}$ at a given site if 
\begin{enumerate}[$\bullet$] 
\item every action it performs at the site is contained in $\pol{T}$ 
\item it will only migrate to sites whose names are in $\pol{T}$. 
\end{enumerate} 
For example, conforming to the policy
\begin{math}
 \sset{\Cinfo,\;\Crequest,\;\Chome} 
\end{math}, 
where \Cinfo,\Crequest are actions and \Chome a location, means that the only 
actions that will be performed are from the set 
\begin{math}
  \sset{\Cinfo,\;\Crequest}
\end{math} 
and migration will only occur, if at all, to the site \Chome. 
With this interpretation of policies, our definition of the predicate $\enforces$ 
is also intuitive; if some code $P$ conforms to the policy $\pol{T}_1$ 
and $\pol{T}_1 \enforces \pol{T}_2$ then $P$ also
automatically conforms to $\pol{T}_2$.

The purpose of membranes is to enforce such policies on incoming agents. 
In other words, at a site $\node l \memb Q$ wishing to enforce a policy 
$\pol{T_{in}}$, the membrane $\memb$ has to decide when to allow entry to 
an agent such as $\gopcc l \pcc P$ from another site. There are two possibilities. 
\begin{enumerate}[$\bullet$]
\item The first is to syntactically check the code $P$ against the policy
   $\pol{T_{in}}$; an implementation would actually expect the agent to
   arrive with a proof of this fact, and this proof would be checked.
\item The second would be to \emph{trust} the agent that its code $P$
   conforms to the stated $\pol{T}$ and therefore only check that this
   conforms to the entry policy $\pol{T_{in}}$.
   Assuming that checking one policy against another is more efficient
   than the code analysis, this would make entry formalities much
   easier. 
\end{enumerate} 
Deciding on when to apply the second possibility presupposes a 
\emph{trust management} framework for systems, which is the topic of
much current research. To simplify matters, here we simply assume that 
each site contains, as part of its membrane, a record of the level of trust 
it has in other sites. Moreover, we assume only three possible levels: 
$\lbad, \loc$ and $\lgood$. Intuitively, a site is $\lgood/\lbad$ 
if it behaves in a reliable/unreliable way, i.e. it does/doesn't properly
calculate digests. On the other hand, a site tagged as $\loc$ can
behave in a non specified way; thus, for the sake of security, it will
be considered as $\lbad$. In a more realistic scenario, it would be
possible to refine $\loc$ to either $\lgood$ or $\lbad$, upon collection
of enough evidence to consider it reliable or not. For the sake of 
simplicity, we do not model this framework here.

\begin{myfigure}
\[
\begin{array}{lrcll} \vspace*{.4cm}
\Rred{act}&
\node l {\memb} {a.P \ | \ Q} & \red{} & \node l {\memb} {P\,|\,Q}
\\
\vspace*{.4cm} 
\Rred{par} & 
\multicolumn{3}{c}{ \hspace*{-.6cm} 
\prooftree N_1 \ \red{} \ N_1' 
\justifies N_1 \parallel N_2 \red{} N_1' \parallel N_2 
\endprooftree } 
\\
\vspace*{.4cm} 
\Rred{struct}&
\multicolumn{3}{c}{ \hspace*{-.5cm} 
\prooftree N \equiv N_1 \quad N_1 \red{} N_1' \quad N_1' \equiv N' 
\justifies N \red{} N' 
\endprooftree } 
\\
\vspace*{.2cm} 
\Rred{mig}&
\multicolumn{3}{l}{ 
\node k {\memb^k} {\gopcc l \pcc P \ | \ Q} \ \parallel \ \node l {\memb^l} R \quad 
   \red{} \quad } 
\\ 
& \multicolumn{3}{l}{ \hspace*{3.5cm}
\node k {\memb^k} {Q} \ \parallel \ \node l {\memb^l} {P\,|\,R}  } & 
\quad\mbox{{\rm if } $\allows {\memb^l} k \pcc P$} 
\end{array} 
\] 
\caption{The reduction relation
\label{fig:reductions}}  

\[ 
\hspace*{-.4cm} 
\begin{array}{lrcllrcl} 
\vspace*{.2cm} 
& \node l \memb {P\ |\ \mmnil} & \equiv & \node l \memb P & 
& N \parallel \emptynet & \equiv & N  
\\ 
\vspace*{.2cm} 
& \node l \memb {P \ | \ Q} & \equiv & \node l \memb {Q \ | \ P} & 
& N_1 \parallel N_2 & \equiv & N_2 \parallel N_1  
\\ 
\vspace*{.2cm} 
& \node l \memb {(P \, | \, Q) \ | \ R} & \equiv & \node l \memb {P \ | \ (Q \, | \, R)} 
\qquad & 
& (N_1 \parallel N_2) \parallel N_3 & \equiv & N_1 \parallel (N_2 \parallel N_3) 
\\ 
& \node l \memb {!P \, | \, Q} & \equiv & \node l \memb {P \, | \, !P \, | \, Q} 
\\ 
\end{array} 
\] 
\caption{The structural equivalence
\label{fig:structeq}}
\end{myfigure}

\begin{defi}[Membranes] A membrane $\memb$ is a pair 
$(\memb_t,\pol{\memb_p})$ where 
\begin{enumerate}[$\bullet$]
\item $\memb_t$ is a partial function from $\Locs$ to $\sset{\loc,\lgood,\lbad}$
\item $\pol{\memb_p}$ is a policy
\end{enumerate} 
\end{defi}

\subsection{The Operational Semantics} \label{op.sem}
\ \\

Having defined both \emph{policies} and \emph{membranes}, we now give an operational
semantics for the calculus, which formalises the above discussion 
on how to manage agent migration.
This is given as a binary relation $N \red{} N'$ over
systems; it is defined to be the least relation which satisfies the rules in
Figure~\ref{fig:reductions}. Rule \Rred{act}
says that the agent $a.P$ running in parallel with other code in site $l$, such as
$Q$, can perform the action $a$; note that the
semantics does not record the occurrence of $a$.
\Rred{par} and \Rred{struct} are standard. The first allows
reductions within parallel components, while the second says that
reductions are relative to a structural equivalence; the rules 
defining this equivalence are given in Figure~\ref{fig:structeq}.
The interesting reduction rule is the last one, \Rred{mig}, governing migration; 
the agent $\gopcc l \pcc P$ can migrate from site $k$ to site $l$ provided 
the predicate $\allows {\memb^l} k \pcc P$ is true. This `enabling' predicate formalises 
our discussion above on the role of the membrane $\memb^l$, and requires in turn
a notion of code $P$ satisfying a policy $\pol{T}$,
\begin{displaymath}
 \types P: \pol{T}
\end{displaymath} 
With such a notion, we can then define $\allows {\memb^l} k \pcc P$ to be: 
\begin{equation}\label{eq:allows}
  \mmif \memb^l_t(k) = \lgood
  \mmthen (\pol{\pcc} \enforces \pol{\memb_p}^{\!\!\!l}\ )
  \mmelse \types P: \pol{\memb_p}^{\!\!\!l} 
\end{equation} 
In other words, if the target site $l$ trusts the source site $k$, 
it trusts that the professed policy $\pol{T}$ is a faithful reflection of the 
behaviour of the incoming agent $P$, and then entry is gained provided that $\pol{T}$ 
enforces the entry policy $\pol{\memb_p}^{\!\!\!l}$\,
(i.e., in this case, $\pol{\pcc} \subseteq \pol{\memb_p}^{\!\!\!l}$\,).
Otherwise, if $k$ can not be trusted, then the entire incoming code $P$ has to be 
checked to ensure that it conforms to the entry policy, as expressed by the predicate 
$\types P : \pol{\memb_p}^{\!\!\!l}$\,.

In Figure~\ref{fig:typechecking1} we describe a simple inference system for 
checking that agents conform to policies, i.e.\ to infer judgements of the form 
$\types P: \pol{T}$. 
Rule \Rtype{empty} simply says that the empty agent
$\mmnil$ satisfies all policies. 
\Rtype{act} is also straightforward; $a.P$ satisfies a policy $\pol{T}$ 
and if $a$ is allowed by $\pol{T}$, and the residual $P$ satisfies $\pol{T}$. 
The rule \Rtype{par} says that to check $P \ | \ Q$ it is sufficient to check $P$ 
and $Q$ separately,
and similarly for replicated agents.
The most interesting rule is \Rtype{mig}, which checks $\gopcc l {\pol{T}'} P$.
This not only checks that migration to $l$ is allowed by the policy, 
that is $l \in \pol{T}$, but it also checks that the code to be spawned there, $P$, 
conforms to the associated professed policy $\pol{T}'$. 
In some sense, if the agent $\gopcc l {\pol{T}'} P$ is allowed  entry into a site $k$, 
then $k$ assumes responsibility for any promises that it makes about conformance to 
policies.

\begin{myfigure}[t] 
\begin{displaymath} 
\begin{array}{lll} \vspace*{.3cm}
 \slinfer[\Rtype{empty}]
       {\types \mmnil:\pol{T}} \qquad &
 \linferSIDE[\Rtype{act}]
      {\types P:\pol{T} }
      {\types a.P:\pol{T}}
      {\!\!\!a \in \pol{T}} &
 \linferSIDE[\Rtype{mig}]
    {\types P:\pol{T}'}
    {\types \gopcc l {\pol{T}'} P: \pol{T}}
    {\!\!l \in \pol{T}} 
\\ 
\linfer[\Rtype{repl}]
    {\types P : \pol T}
    {\types\ !P: \pol{T}} &
 \linfer[\Rtype{par}]
     {\types P:\pol{T} \qquad \types Q:\pol{T}}
     {\types P \ | \ Q:\pol{T}} 
\end{array} 
\end{displaymath}
\caption{Typechecking incoming agents}
\label{fig:typechecking1} 
\end{myfigure}

\subsection{Safety}\label{sect:safety}
\ \\

We have just outlined a reduction semantics in which sites seek to enforce policies 
either by directly checking the code of incoming agents against entry policies, or 
more simply by checking the professed policy of trusted agents. 
The extent to which this strategy works depends, not surprisingly, on the quality 
of a site's trust management.

\begin{exa}\label{ex:one}
Let $\Chome$ be a site name with the following trust function
$$
     \memb^{h}_t \ :\ \{\Calice, \Cbob, \Csecure\} \quad \mapsto\quad \lgood\ .
$$
Consider the system 
\begin{displaymath} 
N \ \define\ \node {\Chome} {\memb^h} {P^h}
 \ \parallel \
 \node {\Cbob} {\memb^b} {P^b}
 \ \parallel \
 \node {\Calice} {\memb^a} {P^a}
 \ \parallel \
 \node {\Csecure} {\memb^s} {P^s} 
\end{displaymath} 
in which the entry policy of $\Chome$, $\pol{\memb_p}^{\!\!\!h}$, is 
$\sset{\Cinfo,\;\Crequest,\;\Csecure}$, and that of $\Csecure$, 
$\pol{\memb_p}^{\!\!\!s}$, is $\sset{\Cgive,\;\Chome}$.
Since $\memb_t^h(\Cbob) = \lgood$, agents migrating from $\Cbob$ to 
$\Chome$ are trusted and only their digests are checked against the entry policy 
$\pol{\memb_p}^{\!\!\!h}$. So, if $P^b$ contains the agent 
\begin{displaymath}
\gopcc \Chome  {\pol{T}_1} {(\Ctake.Q)} 
\end{displaymath} 
where $\pol{T}_1 \enforces \pol{\memb_p}^{\!\!\!h}$, then the entry policy 
of $\Chome$ will be transgressed.

As another example, suppose $\Calice$, again trusted by $\Chome$, contains 
the agent 
\begin{displaymath}
 \gopcc \Chome {\pol{T}_1} {(\Cinfo.\gopcc \Csecure  {\pol{T}_2} {(\Ctake.Q)} )} 
\end{displaymath} 
where $\pol{T}_2$ is some policy which enforces the entry policy of $\Csecure$, 
$\pol{\memb_p}^{\!\!\!s}$. Again because $\pol{T}_1 \enforces 
\pol{\memb_p}^{\!\!\!h}$\,, the migration is  allowed from $\Calice$ to $\Chome$, 
and moreover the incoming agent conforms to the policy demanded of $\Chome$. 
The second migration of the agent is also successful if $\Csecure$ trusts 
$\Chome$: $\memb^s_t(\Chome) = \lgood$ and therefore only the digest 
$\pol{T}_2$ is checked against the entry policy of $\Csecure$. 
We then have the reduction 
\begin{displaymath}
 N \ \redstar{} \ \Chome [\![\ldots]\!]
 \ \parallel \
 \Cbob [\![\ldots]\!]
 \ \parallel \
 \Calice [\![\ldots]\!]
 \ \parallel \
 \node {\Csecure} {\memb^s}{\Ctake.Q \ | \ P^s}
\end{displaymath} 
in which now the entry policy of $\Csecure$ has been foiled. 
\end{exa}

The problem in this example is that the trust knowledge of \!\!\Chome is 
faulty; it trusts in sites which do not properly ensure that professed 
policies are enforced.
Let us divide the sites into \emph{trustworthy} and otherwise. 
This bipartition could be stored in an external record stating which nodes 
are trustworthy (i.e. typechecked) and which ones are not.
However, for economy, we prefer to record this information in the membranes, 
by demanding that the trust knowledge at trustworthy sites is a proper reflection 
of this division. 
This is more easily defined if we assume the following ordering over trust levels:  
\begin{displaymath} 
\loc \dpisubt \lbad \qquad\text{and}\qquad \loc \dpisubt \lgood 
\end{displaymath} 
This reflects the intuitive idea that sites classified as $\loc$ may, 
perhaps with further information, be subsequently classified either as $\lgood$ 
or $\lbad$. On the other hand, $\lgood$ or $\lbad$ cannot 
be further refined; sites classified as either, will not be reclassified.

\begin{defi}[Trustworthy sites and Coherent systems] 
In a system $N$, the site $k$ is \emph{trustworthy} 
if $\memb_t^k(k) = \lgood$.
$N$ is \emph{coherent} if  $\memb_t^k(l) \dpisubt \memb_t^l(l)$
for every trustworthy site $k$. 
\end{defi} 
Thus, if a trustworthy site $k$ believes that a site $l$ can be trusted 
(i.e., $\memb_t^k(l) = \lgood$), then $l$ is indeed trustworthy (as represented by 
$\memb_t^l(l) = \lgood$). Similarly, if it believes $l$ to be $\lbad$, then 
$l$ is indeed bad. The only uncertainty is when $k$ classifies $l$ as $\loc$: 
then $l$ may be either $\lgood$ or $\lbad$. Of course, in coherent systems we 
expect sites which have been classified as trustworthy to act in a 
\emph{trustworthy manner}, which amounts to saying that code running at such 
a $k$ must have at one time gained entry there by satisfying the entry policy.
Note that by using policies as in Definition~\ref{def:policies}, if $P$ 
satisfies an entry policy $\pol{\memb_p}^{\!\!\!k}$, then it continues to satisfy 
the policy while running at $k$ (cf.\ Theorem~\ref{safety} below).

This property of coherent systems, which we call \emph{well-formedness},
can therefore be checked syntactically.
In Figure~\ref{fig:wellformed1}, we give the set of rules
for deriving the judgement 
\begin{displaymath}
 \types N : \tok 
\end{displaymath} 
of well-formedness of $N$.
There are only two interesting rules. Firstly, \Rstype{g.site} 
says that $\node l \memb P$ is well-formed whenever $l$ is trustworthy and
$\types P: \pol{\memb_p}$.
There is a subtlety here; this not only means that $P$ conforms to 
the policy $\pol{\memb_p}$, but also that any digests proffered by agents in $P$
can also be trusted. The second relevant rule is \Rstype{u.site}, 
for typing unknown sites: here there is no need to check the resident code, 
as agents emigrating from such sites will not be trusted.

\begin{myfigure}[t] 
\begin{displaymath}
\begin{array}{ll} \vspace*{.4cm}
 \slinfer[\Rstype{empty}]
     { \types \emptynet:\tok} &
 \linferSIDE[\Rstype{g.site}]
      {\types P:\pol{\memb_p}  }
      {\types \node l \memb P :\tok  }
      {\!\!l \;\;\text{trustworthy}} 
\\
 \linfer[\Rstype{par}]
     { \types N_1:\tok,\qquad \types N_2:\tok}
     {\types N_1 \ \parallel \ N_2:\tok} \qquad &
 \linferSIDE[\Rstype{u.site}]
      {}
      {\types \node l \memb P :\tok}
      {\!\!l \;\;\text{not trustworthy}} 
\end{array} 
\end{displaymath}
\caption{Well-formed systems}
\label{fig:wellformed1} 
\end{myfigure}

\begin{exa}(\textit{Example~\ref{ex:one} continued.})\hskip0.5em 
Let us now re-examine the system $N$ in Example~\ref{ex:one}.
Suppose $\Chome$ is trustworthy, that is $\memb^h_t(\Chome) = \lgood$. 
Then, if $N$ is to be coherent, it is necessary for each of the sites 
$\Cbob$, $\Calice$ and $\Csecure$ also to be trustworthy. Consequently, 
$N$ cannot be well-formed. For example, to derive $\types N: \tok$ it 
would be necessary to derive 
\begin{displaymath}
 \types \gopcc \Chome  {\pol{T}_1} {(\Ctake.Q)}  :\pol{\memb_p}^{\!\!\!b} 
\end{displaymath} 
where $\pol{\memb_p}^{\!\!\!b}$ is the entry policy of $\Cbob$.
But this requires the judgement 
\begin{displaymath}
 \types \Ctake.Q :\pol{T}_1 
\end{displaymath} 
where $\pol{T}_1 \enforces \pol{\memb_p}^{\!\!\!h}$. 
Since $\Ctake \not\in \pol{\memb_p}^{\!\!\!h}$, this is not possible. 

One can also check that the code running at $\Calice$ stops the system from 
being well-formed. Establishing $\types N: \tok$ would also
require the judgement 
\begin{displaymath}
 \types \gopcc \Chome {\pol{T}_1} {(\Cinfo.\gopcc \Csecure
   {\pol{T}_2} {(\Ctake.Q)} )}
  \;\;:\pol{\memb_p}^{\!\!\!a} 
\end{displaymath} 
which in turn, eventually, requires
\begin{displaymath}
 \types \Ctake.Q :\pol{T}_2 
\end{displaymath} 
for some $\pol{T}_2$ such that $\pol{T}_2 \enforces \pol{\memb_p}^{\!\!\!s}$;
this is impossible, again because $\Ctake$ is not in $\pol{\memb_p}^{\!\!\!s}$.
\end{exa}

In well-formed systems we know that entry policies have been respected. 
So one way of demonstrating that our reduction strategy correctly enforces these 
policies is to prove that
\begin{enumerate}[$\bullet$] 
\item system well-formedness is preserved by reduction
\item only legal computations take place within trustworthy sites
\end{enumerate} 
The first requirement is straightforward to formalize: 

\begin{thm}[Subject Reduction] \label{subjRed} 
If $\types N :\tok $ and $N \red{} N'$, then $\types {N'} :\tok$. 
\end{thm} 
\proof
See Appendix~\ref{subjRedOne} \qed

\begin{myfigure}[t] 
\begin{displaymath} 
\begin{array}{llll}
 \slinfer[\Rlts{act}]
    {a.P \redar a P} \quad &
 \slinfer[\Rlts{mig}]
    {\gopcc l {\pol T} P \redar l \mmnil} \quad &
 \linfer[\Rlts{repl}]
    {P\ |\ !P \redar\alpha P'}
    {!P \redar\alpha P'} \quad &
 \linfer[\Rlts{par}]
    {\ \, P_1 \redar\alpha P_1'}
    {\begin{array}{l}
       P_1 \ | \ P_2 \redar\alpha P_1' \ | \ P_2 \\
       P_2 \ | \ P_1 \redar\alpha P_2 \ | \ P_1' \\
     \end{array}} 
\end{array} 
\end{displaymath}
\caption{A Labelled Transition System}
\label{LTS} 
\end{myfigure}

To formalise the second requirement we need some notion of the \emph{computations} 
of an agent. With this in mind, we first define a labelled transition system between 
agents, which details the immediate actions an agent can perform, and the residual 
of those actions. The rules for the judgements 
\begin{displaymath} 
P \redar \alpha Q
\end{displaymath} 
where we let $\alpha$ to range over $\Act \cup \Locs$,
are given in Figure~\ref{LTS}, and are all straightforward. 
These judgements are then extended to
\begin{displaymath} 
P \redar \sigma Q
\end{displaymath} 
where $\sigma$ ranges over
$(\Act \cup \Locs)^*$, in the standard manner:
$\sigma = \alpha_1,\ldots,\alpha_k$, when there exists $P_0,\ldots,P_k$ such that
$P = P_0 \redar{\alpha_1} \ldots \redar{\alpha_k} P_k = P'$. 
Finally, let $\Sset{\sigma}$ denote the set of all
elements of $\Act \cup \Locs$ occurring in $\sigma$.

\begin{thm}[Safety] \label{safety} 
Let $N$ be a well-formed system.
Then, for every trustworthy site $\node l \memb P$ in $N$,
$P \redar\sigma P'$ implies that
$\Sset{\sigma} \enforces \pol {M_p}$. 
\end{thm} 
\proof
See Appendix~\ref{subjRedOne} \qed 

\section{Entry Policies}
\label{entryPol}

The calculus of the previous section is based on a simple notion of entry policies, 
namely finite sets of actions and location names. 
An agent conforms to such a policy $\pol{T}$ at a site if it only executes actions 
in $\pol{T}$ before migrating to some location in $\pol{T}$. 
However both the syntax and the semantics of the calculus are completely 
parametric on policies. All that is required of the collection of policies is
\begin{enumerate}[$\bullet$]
\item a binary relation $\pol{T}_1 \enforces \pol{T}_2$ between them 
\item a binary relation $\types P: \pol{T}$ indicating that the code
      $P$ conforms to the policy $\pol{T}$. 
\end{enumerate} 
With any collection of policies, endowed with two such relations, 
we can define the predicate \linebreak 
$\allows \memb k \pcc P$ as in (\ref{eq:allows}) above, 
and thereby get a reduction semantics for the calculus.
In this section we investigate two variations on the notion of entry policies and 
discuss the extent to which we can prove that the reduction strategy correctly 
implements them.

\subsection{Multisets as Entry Policies} 
\label{multisetsOne}

The policies of the previous section only express the legal actions agents 
may perform at a site.
However in many situations more restrictive policies are desirable. 
To clarify this point, consider the following example.

\begin{exa}\label{ex:two} 
Let $\Cmailserv$ be the site name of a mail server
with the following entry policy $\pol{\memb_p}^{\!\!\!ms}$: 
\begin{eqnarray*}
\sset{\Clist, \Csend, \Cretr, \Cdel, \Creset, \Cquit}
\end{eqnarray*} 
The server accepts client agents performing requests for listing mail messages, 
sending/retrieving/deleting messages, resetting the mailbox and quitting. 
Now, consider the system 
\begin{displaymath} 
N \ \ \define\ \ \node {\Cmailserv} {\memb^{ms}} {P^{ms}}
 \ \ \parallel \
\node {\Cspam} {\memb^s} {\gopcc \Cmailserv \pcc {(!\Csend)}}
\end{displaymath} 
where $\pcc = \{\Csend\}$. According to the typechecking of 
Figure~\ref{fig:typechecking1}, we have that 
$$ 
\types \ !\,\Csend : \pol{\memb_p}^{\!\!\!ms} 
$$ 
However, the agent is a spamming virus and, in practical implementations,
should be rejected by \Cmailserv. 
\end{exa}

\noindent In such scenarios it would be more suitable for policies to be able 
to fix an upper-bound over the number of messages sent. 
This can be achieved in our setting by changing policies from sets of agent 
actions to \emph{multisets} of actions. Consequently, predicate $\enforces$
is now multiset inclusion.

First let us fix some notation.
We can view a multiset as a set equipped with an {\em occurrence function}, 
that associates a natural number to each element of the set.
To model permanent resources, we also allow the occurrence function to 
associate $\omega$ to an element with an infinite number of occurrences 
in the multiset. Notationally, ${\tt e}^\omega$ stands for an element 
${\tt e}$ occurring infinitely many times in a multiset.
This notation is extended to sets and multisets; for any set/multiset $E$, 
we let $E^\omega$ to denote the multiset $\{{\tt e}^\omega : {\tt e} \in E\}$.

\begin{exa}(\textit{Example~\ref{ex:two} continued}.)\hskip0.5em 
\label{ex:two-cont} 
Coming back to Example~\ref{ex:two}, it would be sufficient to define 
$\pol{\memb_p}^{\!\!\!ms}$ to be $\sset{\ldots,\Csend^K,\ldots}$ 
where $K$ is a reasonable constant. In this way, an agent can only send at 
most $K$ messages in each session; if it wants to send more messages, 
it has to disconnect from $\Cmailserv$ (i.e.\ leave it) and then reconnect 
again (i.e.\ immigrate again later on). In practice, 
this would prevent major spamming attacks, because the time spent for 
login/logout operations would radically slow down the spam 
propagation. 
\end{exa}

\begin{myfigure}[t] 
\begin{displaymath}
\begin{array}{lll} \vspace*{.3cm}

\slinfer[\Rtype{empty}]
       {\types \mmnil:\pol{T}} &

\linfer[\Rtype{act}]
       {\types P:\pol{T} }
       {\types a.P:\pol{T \cup \{a\}}} &

\linfer[\Rtype{mig}]
       {\types P:\pol{T}'}
       {\types \gopcc l {\pol{T}'} P: \pol{T} \cup \{l\}} 
\\

\linfer[\Rtype{par}]
       {\types P:\pol{T_1} \qquad \types Q:\pol{T_2}}
       {\types P \ | \ Q:\pol T_1 \cup \pol T_2} \quad &

\linferSIDE[\Rtype{repl}]
       {\types P : \pol T}
       {\types \ !P : \pol{T'}}
       {\!\!\pol T^\omega \enforces \pol{T'}} 
\end{array} 
\end{displaymath}
\caption{Typechecking with policies as Multisets}
\label{fig:typechecking2} 
\end{myfigure}

The theory presented in Sections~\ref{op.sem} and~\ref{sect:safety} 
can be adapted to the case where policies are multisets of actions. 
The judgment $\types P : \pol T$ is redefined in Figure~\ref{fig:typechecking2}, 
where operator $\cup$ stands for \textit{multiset union}. The key rules are 
\Rtype{act}, \Rtype{par} and \Rtype{repl}. The first two properly decrease the 
type satisfied when typechecking sub-agents. The third one is needed because 
recursive agents can be, in general, freely unfolded; hence, the actions they 
intend to locally perform can be iterated arbitrarily many times. For instance, agent 
$$ 
P \ \define \ !\,\Csend 
$$ 
satisfies policy $\pol T \ \define \ \{\Csend^\omega\}$. 
Notice that the new policy satisfaction judgement prevents the spamming virus of 
Example~\ref{ex:two} from typechecking against the
policy of $\Cmailserv$ defined in Example~\ref{ex:two-cont}.

The analysis of the previous section can also be repeated here but 
an appropriate notion of \emph{well-formed} system is more difficult to formulate. 
The basic problem stems from the difference
between \emph{entry} policies and \emph{resident} policies. 
The fact that all agents who have ever entered
a site $l$ respects an entry policy $\pol{\memb_p}$ gives no guarantees as 
to whether the joint effect with the code currently occupying
the site $l$ also satisfies $\pol{\memb_p}$.
For instance, in the terms of Example~\ref{ex:two-cont}, 
$\Cmailserv$ ensures that each incoming agent can only send at most $K$ messages.
Nevertheless, two such agents, having gained entry and now
running concurrently at $\Cmailserv$, can legally send -- jointly -- up to $2K$ messages.
It is therefore necessary to formulate \emph{well-formedness} in terms of 
the individual threads of the code currently executing at a site. 
Let us say $P$ is a \emph{thread} if it is not of the form $P_1 \, |\,P_2$. 
Note that every agent $P$ can be written in the form of $P_1 | \ldots |P_n, n \geq 1$, 
where each $P_i$ is a thread.
So the well-formedness judgment is modified by replacing rule \Rstype{g.site} 
in Figure~\ref{fig:wellformed1} as below. 
\[ 
\linferSIDE[\Rstype{g.site_m}]
    { \forall i \,.\ (P_i \;\text{a thread and }
      \ \types P_i:\pol{\memb_p}) }
    {\types \node l \memb {P_1|\ldots|P_n} :\tok }
    {\!\!l \;\;\rlap{trustworthy}} 
\]

\begin{thm}[Subject Reduction for multiset policies] 
\label{subjRed2} 
If $\types N :\tok $ and $N \red{} N'$, then $\types {N'} :\tok$. 
\end{thm} 
\proof
Similar to that of Theorem~\ref{subjRed}. 
The necessary changes are outlined in
Appendix~\ref{subjRedTwo}.\qed 

The statement of safety must be changed to reflect the focus on 
individual threads rather than agents. Moreover, we must keep into
account also multiple occurrences of actions in a trace; thus,
we let $\Sset{\sigma}$ return a multiset formed by all the actions
occurring in $\sigma$.

\begin{thm}[Safety for multiset policies] 
\label{safetyMulti} 
Let $N$ be a well-formed system. Then, for every
trustworthy site $\node l \memb {P_1|\ldots|P_n}$ in $N$, where each $P_i$ is a thread,
$P_i \redar\sigma P_i'$ implies that
$\Sset{\sigma} \enforces \pol {M_p}$.
\end{thm}
\proof
See Appendix~\ref{subjRedTwo}. \qed

\subsection{Finite Automata as Entry Policies} 
\label{automataOne}

A second limitation of the setting presented in Section~\ref{calculus}
is that policies will sometimes need to prescribe a precise order for 
executing legal actions. This is very common in client/server interactions, 
where a precise protocol (i.e.\ a pattern of message exchange) must be respected. 
To this end we define policies as
{\em deterministic finite automata} (DFAs, for short).

\begin{exa}\label{ex:three}
Let us consider Example~\ref{ex:two} again. Usually, mail servers
requires a preliminary authentication phase to give access to mail services. 
To express this fact, we could implement the entry policy
of $\Cmailserv$, $\pol{\memb_p}^{\!\!\!ms}$, to be the automaton
associated to the regular expression below.
\begin{displaymath}
   \Cusr.\Cpwd.(\Clist + \Csend + \Cretr + \Cdel + \Creset)^*.\Cquit
\end{displaymath} 
The server accepts client requests only upon authentication, via a 
username/password mechanism. Moreover, the policy imposes that each session is 
regularly committed by requiring that each sequence of actions is terminated by 
$\Cquit$. This could be needed to save the status of the transaction and 
avoid inconsistencies. 
\end{exa}

We now give the formal definitions needed to adapt the theory developed in 
Section~\ref{calculus}. We start by defining a DFA, the language associated to it, 
the $\enforces$ predicate between DFAs and a way for an agent to satisfy a DFA. 
As usual \cite{Hopc.Ullm:intro}, a DFA is a quintuple $\pol A \ \define\
(S,\Sigma,s_0,F,\delta)$ where 
\begin{enumerate}[$\bullet$] 
\item $S$ is a finite set of {\em states}; 
\item $\Sigma$ is the {\em input alphabet}; 
\item $s_0 \in S$ is a reserved state, called the {\em starting state}; 
\item $\emptyset \subset F \subseteq S$ is the set of {\em final states}
      (also called {\em accepting states}); 
\item $\delta : S \times \Sigma \to S$ is the
      {\em transition relation}. 
\end{enumerate}
In our framework, the alphabet of the DFAs considered is a finite subset of 
$\Act \cup \Locs$. Moreover, for the sake of simplicity, we shall always
assume that the DFAs in this paper are minimal.

\begin{defi}[DFA Acceptance and Enforcement] 
\label{DFAacp} 
Let $\pol A$ be a DFA. Then 
\begin{enumerate}[$\bullet$]
\item $Acp_s(\pol A)$ contains all the $\sigma \in \Sigma^*$ such that $\sigma$ leads
      $\pol A$ from state $s$ to a final state;
\item $Acp(\pol A)$ is defined to be $Acp_{s_0}(\pol A)$; 
\item $\pol A_1 \enforces \pol A_2$ holds true whenever
      $Acp(\pol A_1) \subseteq Acp(\pol A_2)$. 
\end{enumerate} 
\end{defi} 
 
Notice that, as expected, there is an efficient way to extablish 
$\pol A_1 \enforces \pol A_2$, once given the automata $\pol A_1$ and $\pol A_2$ 
(see Proposition~\ref{enforceEffic} in Appendix~\ref{automataProofs}). 
We now formally describe the language associated to an agent
by  exploiting the notion of {\em concurrent regular expressions} 
(\CRE{}, for short) introduced in \cite{concurRegExprPetri} to model concurrent
processes. For our purposes, the following subset of \CRE{} suffices: 
\[ 
e \quad ::= \quad \epsilon \quad | \quad \alpha 
 \quad | \quad e_1.e_2 \quad | \quad e_1 \odot e_2 \quad | \quad e^\otimes
\] 
$\epsilon$ denotes the empty sequence of characters, 
$\alpha$ ranges over $\Act \cup \Locs$, `.' denotes concatenation, 
$\odot$ is the interleaving (or {\em shuffle}) operator and 
$\,^\otimes$ is its closure. Intuitively, if $e$ represents the language $L$, 
then $e^\otimes$ represents 
$\{\epsilon\}\, \cup\, L\, \cup\, L \odot L\, \cup\, L \odot L \odot L \ldots$.
Given a \CRE{} $e$, the language associated to it, written $lang(e)$, 
can be easily defined; a formal definition is
recalled in Appendix~\ref{automataProofs}. 
Now, given a process $P$, we easily define a \CRE{} associated to it. 
Formally 
$$ 
\begin{array}{rclrcl} 
\vspace*{.2cm} 
\CRE{}(\mmnil) & \define & \epsilon &
\CRE{}(a.P) & \define & a.\CRE{}(P) 
\\
\vspace*{.2cm} 
\CRE{}(\gopcc l {\pol A} P) & \define & l &
\CRE{}(P_1\,|\,P_2) & \define & \CRE{}(P_1) \odot \CRE{}(P_2) 
\\
\CRE{}(!P) & \define & \CRE{}(P)^\otimes \end{array} 
$$

\begin{defi}[DFA Satisfaction]
\label{DFAsatisfaction}
An agent $P$ satisfies the DFA $\pol A$, written $\types P : \pol A$, 
if $lang(\CRE{}(P)) \subseteq Acp(\pol A)$, and $\types Q : \pol
A'$ holds for every subagent of $P$ 
of the form $\gopcc l {\pol A'} Q$.
\end{defi}
In Proposition~\ref{enforceEffic}, we prove that
DFA satisfaction is decidable, although extremely hard to establish.
This substantiate our hypothesis that verifying digests is preferable to
inspecting the full code from the point of view computational
complexity.
We are now ready to state the soundness of this variation. It simply
consists in finding a proper notion of well-formed systems.
As in Section~\ref{multisetsOne}, the entry policy can only express 
properties of single threads, instead of coalitions of threads hosted at a site. 
Thus, we modifiy rule \Rstype{g.site} from
Figure~\ref{fig:wellformed1} to:
\[ 
\linferSIDE[\Rstype{g.site_A}]
    { \forall i \,.\ P_i \mbox{ a thread and }\ \
       \exists s \in S\,.\ lang(\CRE{}(P_i)) \subseteq Acp_s(\pol \memb_p)\  }
    {\types \node l \memb {P_1|\ldots|P_n} :\tok }
    {\!\!l \;\;\rlap{trustworthy}}
\]
This essentially requires that the languages associated to each of the threads in 
$l$ are suffixes of words accepted by $\pol{\memb_p}$ 
(cf.\ Theorem~\ref{safetyDFA} below).
Since this may appear quite weak, it is worth remarking that the 
well-formedness predicate is just a `consistency' check, a way to express that
the agent is in a state from where it will respect the policy of $l$.
The soundness theorems are reported below and are proved 
in Appendix~\ref{automataProofs}. 

\begin{thm}[Subject Reduction for automata policies] 
\label{subjRed3} 
If $\types N :\tok $ and $N \red{} N'$, 
then $\types {N'} :\tok$. 
\end{thm}  

\begin{thm}[Safety for automata policies] 
\label{safetyDFA} 
Let $N$ be a well-formed system. Then, for every
trustworthy site $\node l \memb {P_1|\ldots|P_n}$ in $N$, 
where each $P_i$ is a thread, $\sigma \in lang(\CRE{}(P_i))$ 
implies that there exists some $\sigma'
\in Acp(\pol {M_p})$ such that $\sigma' =
\sigma''\sigma$, for some $\sigma''$. 
\end{thm}

We  conclude this section with two interesting properties 
enforceable by using automata.

\begin{exa}[Lock/Unlock]
We have two actions, $\Clock$ and $\Cunlock$, with the constraint 
that each $\Clock$ must be always followed by an
$\Cunlock$. Let $\Sigma_{\tt l} = \Sigma - \{\Clock\}$ and
$\Sigma_{\tt u} = \Sigma - \{\Clock,\Cunlock\}$. Thus, the desired policy 
(written using a regular expression formalism) is
$$
(\Sigma_{\tt l}^*.(\Clock.\Sigma_{\tt u}^*.\Cunlock)^*)^* 
$$
\end{exa}

\begin{exa}[Secrecy]
Let $\Csecret$ be a secret action; we require that, whenever an agent 
performs $\Csecret$, it cannot migrate anymore (this policy enforces that agents 
having performed $\Csecret$
always remain co-located). Let $\Sigma_{\tt s} = \Sigma - \{\Csecret\}$ and
$\Sigma_{\tt g} = \Sigma - \Locs$;
thus, the desired policy is
$$
\Sigma_{\tt s}^*.(\epsilon + \Csecret.\Sigma_{\tt g}^*)
$$
\end{exa}

\section{Resident Policies}
\label{residentPol}

Here we change the intended interpretation of policies. 
In the previous section a policy dictated the proposed behaviour of an agent 
\emph{prior} to execution in a site, at the point of entry.
This implied that safety in well-formed systems was a thread-wise property 
(see rules \Rstype{g.site_M} and \Rstype{g.site_A}). 
Here we focus on policies which are intended to describe the permitted (coalitional) 
behaviour of agents during execution at a site.
Nevertheless these resident policies are still used to determine whether a new agent 
is allowed access to the site in question; entry will only be permitted if the addition 
of this incoming agent to the code currently executing 
at the site does not violate the policy.

Let us consider an example to illustrate the difference between entry 
and resident policies.

\begin{exa}\label{ex:four}
Let $\Clicence$ be the site name of a server that
makes available $K$ licences to download
and install a software product. The distribution policy is
based on a queue: the first $K$ agents landing in the site
are granted the licence, the following ones are denied.
The policy of the server should be
$\pol{\memb_p}^{\!\!\!s} \ \define\ \sset{\Cgetlic^K}$.
However if this policy is interpreted as an entry policy, applying
the theory of
Section~\ref{multisetsOne}, then the system
grants at most $K$ licences to {\em each} incoming agent. 
Moreover this situation continues indefinitely, effectively
handing out licences to all incoming agents.
\end{exa}

We wish to re-interpret the policies of the previous section as 
\emph{resident policies} and here we outline
two different schemes for enforcing such policies. For simplicity 
we confine our attention to one kind of policy, that of multisets.

\subsection{Static membranes}
\ \\

Our first scheme is conservative in the sense that many of the concepts developed in
Section~\ref{multisetsOne} for entry policies can be redeployed. 
Let us reconsider rule \Rred{mig} from Figure~\ref{fig:reductions}.
There, the membrane $\memb^l$ only takes into consideration
the incoming code $P$, and its digest $\pol{T}$, when deciding
on entry, via the predicate $\allows {\memb^l} k \pcc P$. But if the 
membrane is to enforce a \emph{resident policy}, then it must also take into 
account the contribution of the code already running in $l$, namely $R$. 
To do so we need a mechanism for \emph{joining} policies, such as those 
of the incoming $P$ and the resident $R$ in rule \Rred{mig}. 
So let us assume that the set of policies, with the relation $\enforces$ 
is a partial order in which every pair of elements $\pol{T_1}$ and 
$\pol{T_2}$ has a \emph{least upper bound}, denoted $\pol{T_1 \join T_2}$.
For multiset policies this is the case as $\join$ is simply multiset union. 
In addition we need to be able to calculate the (minimal) policy which a process 
$R$ satisfies; let us denote this as $\Ptype{R}$. For multiset policies 
we can adjust the rules in Figure~\ref{fig:typechecking2}, essentially by 
eliminating \emph{weakening}, to perform this calculation; the resulting rules 
are given in Figure~\ref{fig:typechecking4}, with judgements of the form 
$\infers P:\pol{T}$.\medskip

\begin{lem}~\label{inferenceSound}\hfill
\begin{enumerate}[$\bullet$]
\item For every $P$, there is at most one $\pol{T}$ such that
   $\infers P:\pol{T}$. 
\item $\types P:\pol{T}$ implies that there exists some policy
   $\pol{T'}$ such that $\pol{T'} \enforces \pol{T}$ and
   $\infers P:\pol{T'}$. 
\end{enumerate} 
\end{lem} 
\proof
The first statement is proved by structural induction on $P$; 
the second by induction on the derivation $\types P:\pol{T}$. \qed 

\begin{defi}
Define the partial function $\Ptype{\cdot}$ by
letting $\Ptype{P}$ be the unique policy such that
$\infers P:\pol{T}$, if it exists.
\end{defi}

With these extra concepts we can now change rule \Rred{mig} 
in Figure~\ref{fig:reductions} to take the current resident code 
into account. It is sufficient to change the side condition, from 
$\allows {\memb^l} k \pcc P$ to $\allows {\memb^l,R} k \pcc P$, 
where this latter is defined to be
\begin{displaymath}
  \mmif \memb^l_t(k) = \lgood
  \mmthen (\pcc\! \join \Ptype{R}) \enforces \pol{\memb_p}^{\!\!\!l}\
  \mmelse \types P\mid R: \pol{\memb_p}^{\!\!\!l} 
\end{displaymath} 
Here if only the digest needs to be checked then we compare 
$\pcc\! \join \Ptype{R}$, that is the result of \emph{adding} the digest 
to the policy of the resident code $R$, against the resident policy 
$\pol{\memb_p}^{\!\!\!l}$. 
On the other hand if the source site is untrusted we then need to analyse the 
incoming code in parallel with the resident code $R$.
It should be clear that the theory developed in Section~\ref{multisetsOne} 
is readily adapted to this revised reduction semantics. In particular the 
Subject Reduction and Safety theorems remain true; we spare the reader
the details. However it should also be clear that this approach to enforcing 
resident policies has serious practical drawbacks. An implementation would need to: 
\begin{enumerate} 
\item freeze and retrieve the current content of the site, namely the
  agent $R$; 
\item calculate the minimal policy satisfied by $R$ to be merged with
  $P$'s digest in order to check the predicate $\enforces$, or typecheck
  the composed agent $P\,|\,R$; 
\item reactivate $R$ and, according to the result of the checking phase,
  activate $P$.
\end{enumerate} 
Even if the language were equipped with a `passivation' operator, 
as in \cite{stefani:M-calculus}, the overall operation would still be 
computationally very intensive. Consequently we suggest below another approach.

\begin{myfigure}[t] 
\begin{displaymath} 
\hspace*{-.2cm} 
\begin{array}{lll} 
\vspace*{.3cm} 
\slinfer[\Rinf{empty}]
       {\infers \mmnil:\emptyset} & 
\linfer[\Rinf{act}]
      {\infers P:\pol{T} }
      {\infers a.P:\pol{T} \cup \{a\}} & 
\linferSIDE[\Rinf{mig}]
    {\infers P:\pol{T}'}
    {\infers \gopcc l {\pol T} P: \{l\}}
    {\!\!\!\pol T' \enforces \pol T}
\\ 

\linfer[\Rinf{repl}]
    {\infers P: \pol T}
    {\infers \ !P :\pol T^\omega} \qquad & 
\linfer[\Rinf{par}]
     {\infers P:\pol{T_1} \qquad \infers Q:\pol{T_2}}
     {\infers P \ | \ Q:\pol{T_1 \cup T_2}} 
\end{array} 
\end{displaymath} 
\caption{Type inference for agents with policies as multisets} 
\label{fig:typechecking4} 
\end{myfigure}

\subsection{Dynamic membranes}
\ \\

In the previous approach we have to repeatedly calculate the 
policy of the current resident code each time a new agent requests entry.
Here we allow the policy in
the membrane to ``\emph{decrease},'' in order to reflect the 
resources already allocated to the resident code. So at any 
particular moment in time the policy currently in the membrane records 
what resources \emph{remain}, for any future agents who may wish to enter; 
with the entry of each agent there is a corresponding decrease in the membrane's policy.
Formally we need to change the migration rule rule \Rred{mig}  
to one which not only checks incoming code, or digest, against the membrane's policy, 
but also updates the membrane:
$$
\begin{array}{lll}
\vspace*{.2cm}
\Rred{mig'}\quad&
  \node k {\memb^k} {\gopcc l \pcc P \ | \ Q} \ \parallel \
  \node l {\memb^l} R \ \red{}
\\
& \qquad\qquad\,\node k {\memb^k} {Q} \ \parallel \
      \node l {\widehat\memb^l} {P\,|\,R}
& \ \ \mbox{{\rm if } $\allows {\memb^l} k
        \pcc {P \giveme \widehat\memb^l}$}
\end{array}
$$
where the judgement $\allows {\memb^l} k \pcc {P \giveme \widehat\memb^l}$ is
defined as\medskip
$$
\begin{array}{ll}
\vspace*{.2cm}
 \mmlet\ \pol T' = \smash{\left\{
    \begin{array}{ll}
    \pol T & \mbox{if $\memb^l_t(k) = \lgood $} \\
    \Ptype{P} & \mbox{otherwise}
    \end{array}
    \right\}}
 \ \mmin &
 \pol T'\ \enforces\\
&\phantom{\pol T'\ }\hbox{$\pol{\memb_p}^{\!\!\!l} \ \ \land
  \ \ \pol{M_p}^{\!\!\!l} = \pol{\widehat M_p}^{\,l} \join \pol T'
\ \ \land \ \ M_t^l = \widehat M_t^l$}
\end{array}
$$
First notice that if this migration occurs then the membrane at the
target site changes, from $\pol{M_p}^{\!\!\!l}$ to $\pol{\widehat
M_p}^{\,l}$.  The latter is obtained from the former by eliminating
those resources allocated to the incoming code $P$.  If the source
site, $k$, is deemed to be $\lgood$ this is calculated via the
incoming digest $\pol{T}$; otherwise a direct analysis of the code $P$
is required, to calculate $\Ptype{P}$.

This revised schema is more reasonable from an implementation point of view, 
but its soundness is more difficult to formalise and prove.
As a computation proceeds no permanent record is kept in the system of 
the original resident policies at the individual sites. 
Therefore well-formedness can only be defined relative to an external record of 
what the resident policies were, when the system was initiated. 
For this purpose we use a function $\Theta$, mapping trustworthy sites to policies; 
it is sufficient to record the original polices at these sites as we 
are not interested in the behaviour elsewhere.

\begin{myfigure}[t] 
\begin{displaymath}   
\begin{array}{ll} 
\vspace*{.4cm}

\linferSIDE[\Rstype{g.site}]
      {}
      {\Theta \types \node l \memb P :\tok  }  
      {\!\!\!\!\!\begin{array}{l}
       l \;\;\text{trustworthy} \\
       (\Ptype{P} \join \pol{\memb_p}) \enforces \Theta(l)
      \end{array}
      } \ \  &
 \slinfer[\Rstype{empty}]
       {\Theta \types \emptynet:\tok} 
\\

\linferSIDE[\Rstype{u.site}]
      {}
      {\Theta \types \node l \memb P :\tok  }
      {\!\!l \;\;\text{not trustworthy}} &
\linfer[\Rstype{par}]
     {\Theta \types N_1:\tok,\qquad \Theta \types N_2:\tok}
     {\Theta \types N_1 \ \parallel \ N_2:\tok} 
\end{array} 
\end{displaymath} 
\caption{Well-formed systems under $\Theta$} 
\label{fig:wellformed2} 
\end{myfigure}

Then we can define the notion of well-formed systems, 
relative to such a $\Theta$; this is written as $\Theta \types N :\tok$ 
and the formal definition is given in Table~\ref{fig:wellformed2}.
The crucial rule is \Rstype{g.site}, for trustworthy sites.
If $l$ is such a site then $\node l \memb P$ is well-formed relative to 
the original record $\Theta$ if $\pol{M_p}^{\!\!\!l} \join \Ptype P$ 
guarantees the original resident policy at $l$, namely $\Theta(l)$.

\begin{thm}[Subject Reduction for resident policies] 
\label{subjRed4} 
If $\Theta \types N :\tok $ and $N \red{} N'$, then
$\Theta \types {N'} :\tok$. 
\end{thm} 
\proof
Outlined in Appendix~\ref{ProofsubjRedTwo}. \qed

The introduction of these external records of the original resident policies 
also enables us to give a Safety result. 

\begin{thm}[Safety for resident policies] 
\label{safetyResid} Let $N$ be a well-formed system w.r.t. $\Theta$. 
Then, for every trustworthy site $\node l \memb P$ in $N$,
$P \redar\sigma P'$ implies that
$\Sset{\sigma} \enforces \Theta(l)$. 
\end{thm} 
\proof
See Appendix~\ref{ProofsubjRedTwo}.\qed

\section{Conclusion and Related Work}
\label{conclRelWork}

We have presented a framework to describe distributed computations of
systems involving migrating agents. The activity of agents
entering/running in `good' sites is constrained by a membrane that
implements the layer dedicated to the security of the site. We have
described how membranes can enforce several interesting kind of
policies. The basic theory presented for the simpler case has been
refined and tuned throughout the paper to increase the expressiveness
of the framework. Clearly, any other kind of behavioural specification 
of an agent can be considered a policy. For example, a promising 
direction could be considering logical frameworks (by exploiting model 
checking or proof checkers).

The calculus we have presented is very basic: it is even simpler
than CCS \cite{milner:CCS}, as no synchronization can occur. Clearly,
we did not aim at Turing-completeness, but at a very basic framework
in which to focus on the r\^ole of membranes.
We conjecture that, by suitably advancing the theory presented here,
all the ideas can be lifted to more complex calculi
(including, e.g., synchronization, value passing and/or name restriction).

\medskip
\noindent
{\bf Related Work.}
In the last decade, several calculi for distributed systems with code
mobility have appeared in literature. In particular, structuring a
system as a (flat or hierarchical) collection of named sites
introduced the possibility of dealing with sophisticated concrete
features. For example, sites can be considered as the unity of
\textit{failure} \cite{fournet.gonthier.ea:calculus-mobile,amadio:modelling-mobility},
\textit{mobility} \cite{fournet.gonthier.ea:calculus-mobile,CG00:mobile-ambients}
or \textit{access control} \cite{HenRie-IC02,riely.hennessy:trust-partial,GP03:icalp}.
The present work can be seen as a contribution to the last research
line.

As in  \cite{GP03:icalp}, we have presented a scenario where
membranes can evolve. However, the membranes presented in
Section~\ref{residentPol} only describe `what is left' in the
site. On the other hand, the (dynamically evolving) type of a site in
\cite{GP03:icalp} always constrains the overall behaviour of agents
in the site and it is modified upon acquisition/loss of privileges
through computations.

We borrowed from \cite{riely.hennessy:trust-partial} the notion of
{\em trust} between sites, where agents coming from
trusted sites are accepted without any control. Here, we relaxed this
choice by examining the digest of agents coming from trusted sites.
Moreover, we have a fixed net of trust; we believe that, 
once communication is added to our basic framework, the richer scenario of
\cite{riely.hennessy:trust-partial} (where the partial knowledge of a
site can evolve during its computation) can be recovered.

A related paper is \cite{igarashi.kobayashi:generic-type}, where
authors develop a {\em generic} type system that can be smoothly
instantiated to enforce several properties of the \pic\ (dealing with
arity mismatch in communications, deadlock, race control and
linearity).
They work with one kind of type, and modify the
subtyping relation in order to yield several relevant notions of
safety. The main difference with our approach is that we have
different kind of types (and, thus, different type checking
mechanisms) for all the variations we propose. It would be nice to lift
our work to a more general framework closer to theirs; we leave this
for future work.

Our work is also related to \cite{NR04:typed-static-analysis}. Policies
are described there as deterministic finite automata and constrain
the access to critical sections in a concurrent functional language.
A type and effect system is provided that guarantees adherence of
systems to the policy.
In particular, the sequential behaviour of
each thread is guaranteed to respect the policy, and the
interleavings of the threads' locks to be safe.
But unlike our paper \cite{NR04:typed-static-analysis} has no
code migration, and no explicit distribution; thus, only one
centralised policy is used.

Membranes as filters between the computing body of a site and the
external environment are also considered in
\cite{ferrari.moggi.pugliese.2003,Bou04:membranes,stefani:M-calculus}.
There, membranes are computationally capable objects, and can be
considered as a kind of process. They can evolve and communicate
both with the outer and with the inner part of the associated node,
in order to regulate the life of the node. This differs from our
conception of membranes as simple tools for the verification of
incoming agents.

To conclude, we remark that our understanding of membranes is radically
different from the concept of policies in \cite{ES99}. Indeed, in
{\em loc.\ cit.}, security automata control the execution of agents
running in a site by {\em in-lined monitoring}. This technique
consists of accepting incoming code unconditionally, but blocking
at runtime those actions not abiding the site policy. Clearly, in
order to implement the strategy, the execution of each action must be
filtered by the policy. This contrasts with our approach, where
membranes are `containers' that regulate the interactions between
sites and their environments. The computation taking place within the
site is out of the control of the membrane, which therefore   cannot rely
on in-lined monitoring. 

\section*{Acknowledgement}
The authors wish to acknowledge the reviewers of this paper
for their positive attitude and for their fruitful comments.
Joanna Jedrzejowicz kindly answered some questions on 
regular languages with interleaving and iterated interleaving.

\begin{small}

\end{small}

\appendix
\section{Technical Proofs}
\label{techProofs}
We now outline the proofs of the technical results in
the paper, section by section.

\subsection{Proofs of Section~\ref{calculus}}\label{subjRedOne}
\ \\

\begin{lem}[Subsumption]
\label{subtyping}
If $\types P :\pol{T}$ and
$\pol T \enforces \pol{T'}$,
then $\types P :\pol{T'}$.
\end{lem}
\proof
By induction on the derivation of the judgment
$\types P :\pol{T}$.\qed

\paragraph{\bf Proof of Theorem~\ref{subjRed} [Subject Reduction]:} 
The proof is by induction over the inference of $N \red{} N'$. 
Notice that trustworthiness is invariant under  reduction. Therefore
coherence, which is defined in terms of the trustworthiness of sites,
is also  preserved by reduction.

We outline the proof when the inference is deduced using 
rule \Rred{mig}, a typical example.
By hypothesis, $\types \node k {\memb^k} {\gopcc l \pcc P \ | \ Q}
:\tok$; this implies that $\types \node k {\memb^k} Q :\tok$.
Thus, we only need to prove that $\types \node l {\memb^l} R :\tok$
and $\allows {\memb^l} k \pcc P$ imply 
$\types \node l {\memb^l} {P\,|\,R} :\tok$.
We have two possible situations:
\begin{enumerate}[$l$]
\item {\bf trustworthy:}\  
Judgment $\types R : \pol{\memb_p}^{\!\!\!l}\ $ 
holds by hypothesis; 
judgment $\types P : \pol{\memb_p}^{\!\!\!l}\ $ 
is implied by $\allows {\memb^l} k \pcc P$. Indeed, because of the coherence
hypothesis, $M_t^l(k) \dpisubt M_t^k(k)$. If $M_t^k(k) \neq \lgood$, then
$\allows {\memb^l} k \pcc P$ is exactly the required
$\types P : \pol{\memb_p}^{\!\!\!l}\,$. Otherwise, we know that 
$\types \gopcc l \pcc P : \pol{\memb_p}^{\!\!\!k}\,$;
by rule \Rtype{mig} this implies that $\types P : \pcc$.
Judgment $\types P : \pol{\memb_p}^{\!\!\!l}\ $ 
is obtained by using Lemma~\ref{subtyping}, since
$\allows {\memb^l} k \pcc P$ is defined to be
$\pcc \enforces \pol{\memb_p}^{\!\!\!l}\,$
(see (\ref{eq:allows}) in Section~\ref{op.sem}).
Thus, by using \Rtype{par}, we obtain the desired $\types P|R : \pol{M_p}^{\!\!\!l}$\,.
\item {\bf not trustworthy:} This case is simple, because
rule \Rstype{u.site} always allows to derive \linebreak
$\types \node l {\memb^l} {P\,|\,R} :\tok$.
\end{enumerate}
The case when \Rred{act} is used is similar, although simpler, and 
the case when rule \Rred{par} is used requires a simple
inductive argument. Finally to prove the case
when rule \Rred{struct} is used, we need
to know that \emph{coherency} of systems is preserved by 
structual equivalence; the proof of this fact, which is
straightforward, is left to the reader.\qed

\paragraph{\bf Proof of Theorem~\ref{safety} [Safety]:}
Let $\node l \memb P$ be a site in $N$
such that $P \redar\sigma P'$. We have to prove that
$\Sset{\sigma} \enforces \pol {M_p}$.
The statement is proved by induction over the length of $\sigma$.
The base case, when $\sigma = \epsilon$, is trivial since
$\Sset{\epsilon} = \emptyset$.

So we may assume $\sigma = \alpha\sigma'$ and $P \redar{\alpha} P'' 
\redar{\sigma'} P'$. Let us consider $P \redar{\alpha} P''$; 
by induction on $\redar\alpha$, we can prove that $\alpha \in \pol {M_p}$ and that
$\types \node l \memb {P''} : \tok$.
If the transition has been inferred by using rule \Rlts{act},
then $P = a.P''$ and, by rule \Rstype{g.site}, we have that 
$\types a.P'': \pol {M_p}$; by definition of rule \Rtype{act}, 
we have the desired $a \in \pol {M_p}$ and $\types P'': \pol {M_p}$. 
When \Rlts{mig} is used the argument is similar, and all other cases
follow in a straightforward manner by induction.

Thus, we can now apply induction on the number of actions performed in
$P'' \redar{\sigma'} P'$ and obtain
that $\Sset{\sigma'} \enforces \pol {M_p}$. This sufficies to conclude that
$\Sset{\sigma} = (\Sset{\sigma'} \cup \{\alpha\}) \enforces \pol {M_p}$.\qed

\subsection{Proofs of Section~\ref{multisetsOne}}\label{subjRedTwo}
\ \\

The proofs given in Appendix~\ref{subjRedOne} can be easily
adapted  to the setting in which entry policies are multisets.
We outline only  the main changes.
First recall that $\enforces$ is multiset inclusion, that 
judgments $\types P : \pol T$ must be now
inferred by using the rules in Figure~\ref{fig:typechecking2} and 
that rule \Rstype{g.site_M} is used for well-formedness.
Then, Lemma~\ref{subtyping} remains true in this revised setting.

\paragraph{\bf Proof of Theorem~\ref{subjRed2} [Subject Reduction]:}
A straightforward adaptation of 
the corresponding proof in the previous section. 
The only significant change is to the case
when a replication is unfolded via the rule \Rred{struct}, 
i.e.
$$
N \ \ \define \ \
\node l \memb {!P\ | \ Q} \ \ \equiv \ \
\node l \memb {P\ | \ !P\ | \ Q} 
\red{} N'' \ \equiv \ \ N'
$$
By hypothesis, $\types\ !P : \pol{M_p}$; therefore,
by definition of rule \Rtype{repl}, we have that
$\types P : \pol T$ for some $\pol T$ such that $\pol T^\omega \enforces \pol{M_p}$.
Since $\pol T \enforces \pol T^\omega$ and because of
Lemma~\ref{subtyping}, we have that
$\types \node l \memb {P\ | \ !P\ | \ Q} : \tok$.
By induction, $\types N'' : \tok$. It is easy to prove
that this sufficies to obtain the desired $\types N' : \tok$. \qed
 
\paragraph{\bf Proof of Theorem~\ref{safetyMulti} [Safety]:}
From the rule \Rstype{g.site_M} we know that  $\types P_i : \pol {M_p}$,
for all $i=1,\ldots,n$. We now  proceed by induction over $|\sigma|$. The base
case is trivial.
For the inductive case, we consider $\sigma = \alpha\sigma'$
and $P_i \redar\alpha P_i'' \redar{\sigma'} P_i'$. 
By induction on $\redar\alpha$, we can prove that 
$\alpha \in \pol {M_p}$ and that
$\types \node l {\memb_t;\pol {\memb_p} -\{\alpha\}} {P_i''}$.
If the transition has been inferred by using rule \Rlts{act},
then $P_i = a.P_i''$ and, by rule \Rstype{g.site_M}, we have that 
$\types a.P_i'': \pol {M_p}$; by definition of rule \Rtype{act}, 
we have the desired $\pol {M_p} = \pol T \cup \{a\}$ and $\types P'': \pol T$. 
When \Rlts{mig} is used the case is simpler, and all other cases
follow in a straightforward manner by induction.

Coming back to the main claim, we use the induction
and obtain that $\Sset{\sigma'} \enforces \pol {M_p} - \{\alpha\}$;
thus, $\Sset\sigma \enforces \pol {M_p}$.\qed

\subsection{Proofs of Section~\ref{automataOne}}\label{automataProofs}
\ \\

We start by recalling from \cite{concurRegExprPetri}
the formal definition of the language associated to a CRE, 
as follows.
$$
\begin{array}{rcl}
\vspace*{.2cm}
lang(\epsilon) & \define & \{\epsilon\}
\\

\vspace*{.2cm}
lang(\alpha) & \define & \{\alpha\}
\\

\vspace*{.2cm}
lang(e_1.e_2) & \define & \{x_1 x_2 \ : 
        \ x_1 \in lang(e_1) \ \wedge\ x_2 \in lang(e_2)\}
\\

\vspace*{.2cm}
lang(e_1 \odot e_2) & \define & \{x_1 y_1 \cdots x_n y_n \ : 
        \ x_1\cdots x_n \in lang(e_1) \ \wedge\ y_1 \cdots y_n \in lang(e_2)\}
\\

lang(e^\otimes) & \define & \bigcup_{i \geq 0} lang(e)^{\otimes i}
\quad \mbox{where } L^{\otimes i} \ \ \define \ \ \left\{
        \begin{array}{ll}
        \{\epsilon\} & \mbox{if $i = 0$}\\
        L^{\otimes i-1} \odot L & \mbox{otherwise}
        \end{array}
        \right.
\end{array}
$$
Notice that the definition of the $lang(e_1 \odot e_2)$ hides a trick:
the $x_i$s and the $y_i$s can also be $\epsilon$. Thus, as expected,
we can also consider for interleaving strings of different length.

\medskip

We start by accounting on the complexity of predicate $\enforces$ and the 
satisfiability relation when policies are automata. This is stated by the
following Proposition.

\begin{prop}\label{enforceEffic}\hfill
\begin{enumerate}[(1)]
\item $\pol A_1 \enforces \pol A_2$ can be calculated in polynomial time
\item $\types P : \pol A$ is decidable, but it is super-exponential
\end{enumerate}
\end{prop}
\vfill\eject

\proof\hfill
\begin{enumerate}[(1)]

\item Let $\pol A_i = (S_i,\Sigma,s_0^i,F_i,\delta_i)$ and let
$L_i = Acp(\pol A_i)$. By definition, we have to check
whether $L_1 \subseteq L_2$ or not. This is equivalent to check
whether $L_1 \cap \overline L_2 = \emptyset$. 
The following steps have been carried out by following \cite{Hopc.Ullm:intro}.
\begin{enumerate}
\item calculate the automaton associated to $\overline L_2$.
        This can be done in $O(|S_2|)$ and the resulting automaton has
        $|S_2|$ states.
\item calculate the automaton associated to $L_1 \cap \overline L_2$.
        This can be done in $O(|S_1| \times |S_2| \times |\Sigma|)$ and creates 
        an automaton $\pol A$ with $|S_1| \times |S_2|$ states.
\item Checking the emptyness of $L_1 \cap \overline L_2$ can be done by 
        using a breath-first search that starts from the starting state of 
        (the graph underlying) $\pol A$ and stops whenever a final state is reached.
        If no final state is reached, $L_1 \cap \overline L_2$ is empty. This can be done in
        $O(|S_1| \times |S_2| \times |\Sigma|)$.
\end{enumerate}
Thus, the overall complexity is $O(|S_1| \times |S_2| \times |\Sigma|)$.

\item It has been proved in \cite{concurRegExprPetri} that each CRE $e$
can be represented by a (labelled) Petri net, in that the language
accepted by the Petri net is $lang(e)$. Now, we can easily construct
a DFA accepting the complement of the language accepted by $\pol A$
(see item (a) of the previous proof). Now, we can construct the product
between this DFA (that can be seen as a Petri net) 
and the Petri net associated to $CRE(P)$; this Petri net accepts 
$lang(CRE(P)) \cap \overline{Acp(\pol A)}$ (see \cite{Peterson:PetriNetTheory}). 
Now, the emptyness of this language can be solved
with the algorithm for the reachability problem in the corresponding Petri net.
This problem has been proved decidable \cite{reachabilityPetri}
and solvable in double-exponential time \cite{primRecAlgorPetriNets}.\qed
\end{enumerate}

We now prove the subject reduction theorem in the setting where types 
are DFAs. To this aim, we need to adapt Lemma~\ref{subtyping} and we need
a very simple result on the languages associated to DFAs and processes.

\begin{lem}
\label{subtypingDFA}
If $\types P : \pol{A}$ and $\pol A \enforces \pol A'$,
then $\types P :\pol A'$.
\end{lem}
\proof
By transitivity of subset inclusion.\qed

\begin{lem}
\label{langProp}\
\begin{enumerate}
\item $\alpha\sigma \in Acp_s(\pol A)$ if and only if 
        $\sigma \in Acp_{\delta(s,\alpha)}(\pol A)$
\item If $\sigma \in lang(CRE(a.P))$ then $\sigma = a\sigma'$
        for $\sigma' \in lang(CRE(P))$. Viceversa, 
        if $\sigma \in lang(CRE(P))$, then 
        $a\sigma \in lang(CRE(a.P))$.
\end{enumerate}
\end{lem}
\proof
Trivial.\qed

\medskip

\paragraph{\bf Proof of Theorem~\ref{subjRed2} [Subject Reduction]:}
Now $\types N :\tok$ relies on rule
\Rstype{g.site_A}. Again, the proof is by induction
on the inference of $N \red{} N'$. We only give the
base cases, because inductive steps can be handled
with in a standard way. We only consider the cases
of trustworthy sites; the case for non-trustworthy 
sites is easier. In what follows, we write
$\types_s P : \pol A$ to mean that $\types P : \pol A'$,
where $\pol A'$ is the DFA obtained from $\pol A$ by
setting $s$ as starting state.
\begin{enumerate}[$\Rulef{(r\textrm{-}}$]
\item$\!\!\Rulef{act})$ In this case, $N = \node l {\memb} {a.P \ | \ Q}$.
By definition of rule \Rstype{g.site_A}, it holds that $Q = Q_1 | \ldots | Q_k$
(for $Q_i$ threads),
$\exists s.\types_s a.P : \pol{M_p}$ and 
$\forall i.\exists s_i.\types_{s_i} Q_i : \pol{M_p}$.
By definition, we have that $lang(CRE(a.P)) \subseteq Acp_s(\pol{M_p})$;
by Lemma~\ref{langProp}, we have that 
$lang(CRE(P)) \subseteq Acp_{\delta(s,a)}(\pol{M_p})$.
This sufficies to infer the well-formedness of
$N' = \node l {\memb} {P \ | \ Q}$.

\item$\!\!\Rulef{mig})$ In this case, $N = \node k {\memb^k} 
{\gopcc l {\pol A} P \ | \ Q} \ \parallel \ \node l {\memb^l} R$
and $\allows {\memb^l} k {\pol A} P$. We further identify two sub-cases:
\begin{enumerate}[$\bullet$]
\item $M^l(k) = \lgood$. In this case, because of coherence, we know
that $\types P : \pol A$. Moreover, by definition of 
$\allows {\memb^l} k {\pol A} P$, it holds that $\pol A \enforces 
\pol{M_p}^{\!\!\!l}$\,. By Lemma~\ref{subtypingDFA}, we have that
$\types P : \pol{M_p}^{\!\!\!l}$\,. This sufficies to conclude.
\item $M^l(k) \neq \lgood$. This case is simpler because
$\allows {\memb^l} k {\pol A} P$ is defined to be
$\types P : \pol{M_p}^{\!\!\!l}$\,.\qed
\end{enumerate}
\end{enumerate}

\paragraph{\bf Proof of Theorem~\ref{safetyDFA} [Safety]:}
The proof is quite easy. Indeed, by rule \Rstype{g.site_A},
it holds that $\exists s_i : \ \ \types_{s_i} P_i : \pol {\memb_p}^{\!\!\!l}$\,.
By definition, this implies that every $\sigma \in lang(CRE(P_i))$ is in
$Acp_{s_i}(\pol {\memb_p}^{\!\!\!l}\,)$.
Since the automaton $\pol {\memb_p}^{\!\!\!l}\,$ is minimal, $s_i$ is
a reachable state from the starting state $s_0$, say, with a (finite)
string $\sigma''$. Then, by Definition~\ref{DFAacp} and by Lemma~\ref{langProp}.1, 
it holds that $\sigma''\sigma \in Acp(\pol {\memb_p}^{\!\!\!l}\,)$. 
This proves the thesis.\qed

\subsection{Proofs of Section~\ref{residentPol}}\label{ProofsubjRedTwo}
\ \\

We show here the main things to modify to carry out the proofs
given in Appendix~\ref{subjRedTwo}.
Obviously, judgment $\types P : \pol T$
must be now replaced everywhere with $\infers P : \pol T$
and, similarly, $\types N : \tok$ becomes $\Theta \types N : \tok$.

\paragraph{\bf Proof of Theorem~\ref{subjRed4} [Subject Reduction]:}
The proof is by induction over the inference of $N \red{} N'$. 
Inductive steps are simple; we only give the base steps.
\begin{enumerate}[$\Rulef{(r\textrm{-}}$]
\item$\!\!\Rulef{act})$ By hypothesis, $\Theta \types \node l \memb {a.P \ | \ Q}
:\tok$. If $l$ is not trustworthy, the case is trivial. Otherwise, we
know by hypothesis that $(\Ptype{a.P \ | \ Q} \join \pol{M_p}) \enforces \Theta(l)$.
Now, by definition of judgment $\infers$ (and hence of function $\Ptype\cdot$)
we have that $\Ptype{a.P \ | \ Q} = \Ptype{P \ | \ Q} \cup \{a\}$. Hence,
$(\Ptype{P \ | \ Q} \join \pol{M_p}) \enforces \Theta(l)$, as required.

\item$\!\!\Rulef{mig})$ By hypothesis, $\Theta \types \node l {\memb^l} R :\tok$;
we only consider the case in which $l$ is trustworthy. Thus, we know that
$(\Ptype R \join \pol{M_p}^{\!\!\!l}) \enforces \Theta(l)$.
By the premise of rule \Rred{mig}, it holds that
$\allows {\memb^l} k \pcc P \giveme \widehat\memb^l$.
We have two possible situations:
\begin{description}
\item[$M_t^l(k) = \lgood$] 
In this case, $\allows {\memb^l} k \pcc P \giveme \widehat\memb^l$
is defined to be $\pol T \enforces \pol{\memb_p}^{\!\!\!l} \ \wedge\
\pol{M_p}^{\!\!\!l} = \pol{\widehat M_p}^{\,l} \join \pol T \ \wedge
\ M_t^l = \widehat M_t^l$. The fact that $M_t^l = \widehat M_t^l$
is sufficient to preserve coherence. Moreover, by rule \Rinf{mig},
we know that $\infers P:\pol{T}'$ and $\pol T' \enforces \pol T$.
By rule \Rinf{par}, $\Ptype{P|R} = \Ptype R \join \pol T'$ and
$(\Ptype R \join \pol T') \enforces (\Ptype R \join \pol T)$. 
Then, $\Ptype{P|R} \join \pol{\widehat M_p}^l
= (\Ptype R \join \pol T' \join \pol{\widehat M_p}^l) \enforces 
(\Ptype R \join \pol T \join \pol{\widehat M_p}^l) = 
(\Ptype R \join \pol{M_p}^{\!\!\!l}) \enforces \Theta(l)$, as required.

\item[$M_t^l(k) \neq \lgood$] In this case, the previous proof 
should be rephrased by using $\Ptype P$ instead of the digest $\pol T$.\qed 
\end{description}
\end{enumerate}

\paragraph{\bf Proof of Theorem~\ref{safetyResid} [Safety]:}
We prove a slightly more general result, that easily implies
the claim desired.
\begin{quote}
Let $N$ be a well-formed system w.r.t. $\Theta$. 
If $\node l \memb P$ is a trustworthy site of $N$ 
such that $(\Ptype P \join \pol{M_p}^{\!\!\!l}\,) = \pol T$,
then $P \redar\sigma P'$ implies that
$\Sset{\sigma} \enforces \pol T$. 
\end{quote}
The proof is 
by induction over $|\sigma|$. 
The base case is when $\sigma = \epsilon$ and it is trivial.
In the inductive case, we consider $\sigma = \alpha\sigma'$
and $P \redar\alpha P'' \redar{\sigma'} P'$. To start,
it is easy to prove that 
\begin{equation}
\label{EqINCL}
\Ptype{P''} \join \{\alpha\} = \Ptype P 
\end{equation}
By transitivity of multiset inclusion and by the claim (\ref{EqINCL}) above, 
$(\Ptype{P''} \join \pol{M_p}^{\!\!\!l}\,) = \pol {T'}$, where
$\pol T = \pol {T'} \join \{\alpha\}$. 
Thus, node $\node l {\memb^l} {P''}$ is well-formed (and trustworthy). 
By induction we therefore have that $\Sset{\sigma'}
\enforces \pol {T'}$. Hence, $\Sset\sigma = \Sset{\sigma'} \join \{\alpha\}
\enforces \pol{T'} \join \{\alpha\} = \pol T$, as required.

To conclude, the original claim of Theorem~\ref{safetyResid}
is obtained from the result just proved by noticing that,
because of well-formedness, $\pol T \enforces \Theta(l)$.\qed

\end{document}
